\documentclass[11pt]{article}

\usepackage[margin=1in]{geometry}
\usepackage{amsmath,amssymb,bm}
\usepackage{graphicx}
\usepackage{booktabs}
\usepackage{xcolor}
\usepackage{float}          
\usepackage{tikz}
\usetikzlibrary{shapes.geometric, arrows.meta, positioning, fit, backgrounds}
\usepackage[affil-it]{authblk}
\usepackage[numbers,sort&compress]{natbib}
\usepackage[colorlinks=true,linkcolor=blue,citecolor=blue,urlcolor=blue]{hyperref}

\newfloat{algorithm}{htbp}{loa}
\floatname{algorithm}{Algorithm}

\newcommand{\kw}[1]{\textbf{#1}}
\newcommand{\norm}[1]{\left\lVert #1 \right\rVert}
\newcommand{\Pnorm}[1]{\left\lVert #1 \right\rVert_{P}}
\newcommand{\dold}{d_{n}}   
\newcommand{\Aset}{\mathcal{A}}
\newcommand{\Iset}{\mathcal{I}}
\newcommand{\Tr}{^{\mathsf{T}}}
\newcommand{\pred}{\mathrm{pred}}
\newcommand{\ared}{\mathrm{ared}}

\begin{document}
\sloppy

\title{Assessing Nonlinear Elimination Preconditioning for Trust-Region
Phase-Field Fracture}

\author{Tianchen Hu\thanks{\texttt{thu@anl.gov}}}
\affil{Argonne National Laboratory, 9700 S Cass Ave, Lemont, IL 60439, USA}
\date{}

\maketitle

\begin{abstract}
Each quasi-static load step of phase-field fracture is a bound-constrained minimization of a nonconvex,
coupled displacement--damage energy under an irreversibility bound on the damage. Monolithic Newton stalls
once the nonlinearity localizes at the advancing crack front, and staggered (alternate-minimization) schemes
converge slowly there. We present an on-demand nonlinear-elimination preconditioned trust-region Newton
method: an energy Steihaug--Toint trust region, a primal--dual active set for irreversibility, and a
bound-constrained field-split sweep that eliminates an algebraically-identified ``hard set'' spanning both
fields before each step. The elimination is applied \emph{on demand} --- triggered by the coupled Newton's
own stalling and otherwise skipped --- so the method reduces to monolithic Newton at no surcharge where the
step is already healthy. We find the robustness to come from the energy trust region: with that
globalization fixed, monolithic Newton already completes every loading history without cutbacks, where
residual-merit Newton death-spirals, alternate minimization stalls, and the full-field sweep loses
robustness. Against that well-globalized baseline, the on-demand elimination cuts outer nonlinear iterations
by $19$--$25\%$ (brittle) and $17\%$ (ductile), with always-on elimination reaching $26$--$28\%$ and about
$39\%$ at the ductile nucleation step. Measured machine-independently, as a full-mesh-equivalent
assembly-work proxy rather than wall-clock, it is competitive with --- not faster than --- monolithic Newton
(within about $10\%$), whereas an always-on sweep adds up to $30\%$. Nonlinear elimination is thus an
iteration-reduction mechanism whose overhead the on-demand gate bounds, with no demonstrated total-work
advantage over well-globalized monolithic Newton.
\end{abstract}


\medskip

\section{Introduction}\label{sec:intro}

Variational phase-field models of brittle fracture~\citep{FrancfortMarigo1998,BourdinFrancfortMarigo2000,BourdinFrancfortMarigo2008}
regularize a sharp crack by a smooth damage field $d\in[0,1]$ and recast crack evolution as the
minimization of a coupled elastic--fracture energy $\Psi(u,d)$ over displacement $u$ and damage $d$.
Common regularizations (AT1, AT2)~\citep{PhamAmorMarigoMaurini2011,TanneEtAl2018}
differ in whether damage has a purely elastic threshold. Physical irreversibility --- cracks may not
heal --- imposes the bound $d\ge\dold$ at each load step, turning every step into a
\emph{bound-constrained} minimization~\citep{GerasimovDeLorenzis2019,MieheWelschingerHofacker2010}.

Two difficulties make this hard to solve efficiently. First, the energy is nonconvex and the coupled
Hessian becomes indefinite in the softening regime, so a plain (monolithic) Newton iteration on the
coupled system develops a long plateau in the residual once the nonlinearity concentrates at the
advancing crack front. Second, the standard remedy --- \emph{staggered} or alternate-minimization
solvers that alternate between the (convex) displacement and damage subproblems --- is robust but
converges slowly, often requiring hundreds of sweeps per step near
propagation~\citep{BourdinFrancfortMarigo2000,MieheHofackerWelschinger2010,GerasimovDeLorenzis2016}.

\emph{Nonlinear preconditioning}~\citep{BruneKnepleySmithTu2015} attacks exactly this
kind of ``unbalanced'' nonlinearity by transforming the nonlinear system before applying Newton.
Additive and multiplicative Schwarz preconditioned inexact Newton
(ASPIN~\citep{CaiKeyes2002}, MSPIN~\citep{LiuKeyes2015}) precondition by subdomain or by
field split; nonlinear elimination (NEPIN and related
schemes)~\citep{LanzkronRoseWilkes1996,CaiLi2011,LiuEtAl2022}
instead precondition only a small set of ``bad'' (strongly nonlinear) degrees of freedom, eliminating
them before the global Newton step. For fracture, the strongly-nonlinear set is physically the crack
front --- a thin, moving region --- which makes nonlinear elimination a natural fit. Field-split SPIN
has, in fact, already been brought to monolithic phase-field fracture: \citet{KopanicakovaKothariKrause2023}
developed additive and multiplicative field-split (Schwarz) preconditioned inexact Newton for the coupled
displacement--damage system and reported large reductions in nonlinear iterations and wall-clock time over
alternate minimization. That study is the closest precedent to ours, and it is precisely what prompts the
question this paper sets out to answer.

\emph{Globalization} is the second design choice, and it turns out to be as consequential as the
direction. Since the residual is the energy gradient, $R=\nabla\Psi$, the natural merit for this
minimization is $\Psi$ itself; a common alternative instead drives the residual norm $\tfrac12\norm{R}^2$
by a line search. The two coincide only at a nondegenerate equilibrium ($R=0$): the steepest-descent
direction of $\tfrac12\norm{R}^2$ is $-\nabla^2\Psi\,R$, whereas that of $\Psi$ is $-R$. When
$\nabla^2\Psi$ is SPD, $-\nabla^2\Psi\,R$ is also a descent direction for $\Psi$ (since
$R\Tr\nabla^2\Psi\,R>0$); when $\nabla^2\Psi$ becomes indefinite in the softening regime it need not be,
and can point uphill in $\Psi$, so a residual line search then makes little headway on the energy and
stalls near propagation.
$\Psi$ is therefore globalized directly with a Steihaug--Toint trust region, whose truncated CG absorbs the
indefinite Hessian through explicit negative-curvature handling, rather than a line search along a
possibly non-descent Newton direction. Section~\ref{sec:results} bears this out: under the residual merit
the monolithic solve breaks down and the field-split sweep becomes impractically slow through
propagation, while the energy trust region completes without cutbacks.

This second axis also frames how we position the present method against its closest precedent.
\citet{KopanicakovaKothariKrause2023} hold the globalization fixed --- a residual-merit backtracking line
search --- and benchmark monolithic field-split SPIN against alternate minimization; a globalized monolithic
Newton \emph{without} the nonlinear preconditioner is not among the baselines, so the two monolithic
ingredients, the preconditioner and the globalization, are never varied independently. The reported speedup
therefore reflects monolithic coupling, nonlinear preconditioning, and that particular globalization,
\emph{together}, measured against a staggered solver; by construction it cannot isolate the preconditioner's
marginal contribution. This is not a flaw in that study's conclusion --- field-split SPIN is plainly faster
than alternate minimization --- but a question its design leaves open, and it is the one we pursue. Holding
the globalization fixed at the energy trust region we find necessary regardless (the residual merit above
stalls in the softening regime), we ask how much the nonlinear elimination \emph{itself} adds: we therefore
benchmark against a well-globalized monolithic Newton and report cost in machine-independent, full-mesh-%
equivalent assembly-work counts rather than wall-clock time. Measured this way (Section~\ref{sec:results}), the
elimination reduces nonlinear iterations but not overall assembly work; its marginal value over
well-globalized monolithic Newton is therefore iteration reduction, not speed or robustness.

\paragraph{Scope: a variational requirement} This globalization is also the method's main modeling
restriction. Driving the outer iteration by the energy $\Psi$ presumes that the residual is a gradient,
$R=\nabla\Psi$ --- that the model is genuinely \emph{variational}. Several widely-used phase-field variants
deliberately give up this structure to correct the model's strength and nucleation response: the
\emph{hybrid} formulation, which drives damage with the tensile energy but takes the momentum balance from a
differently-degraded stress, so the two equations do not derive from a single energy~\citep{Ambati2015Hybrid};
and the \emph{complete} nucleation model, which adds an external driving force to the damage equation so as to
reproduce a prescribed material-strength surface~\citep{Kumar2020Nucleation}. In both, $R\ne\nabla\Psi$, so no
energy merit exists and the Steihaug--Toint outer loop as formulated here does not apply. The rest of the
construction is unaffected --- the coupled field-split elimination, the subdomain-restricted assembly, and the
primal--dual active set all act on the residual and the bounds directly, independently of any energy --- but
the globalization would have to fall back to the residual merit $\tfrac12\norm{R}^2$, which
Section~\ref{sec:results} finds fragile in exactly the softening regime of interest. Variational routes to
strength are, however, emerging, though they differ in exactly the structure this solver needs. The
complete model has been recast as a \emph{two-functional} (blockwise) minimization~\citep{Larsen2024Variational};
such a formulation restores a variational principle but does not, by itself, furnish a single coupled
potential $\Psi(u,d)$ with $R=\nabla\Psi$ and symmetric $\nabla^2\Psi$ --- the property the Steihaug--Toint
step relies on. Closer to what is required is a fully variational cohesive model with an independently
tunable strength surface~\citep{Vicentini2026Variational}: a \emph{single-potential} formulation that
re-exposes precisely that structure. For readers who need strength-driven nucleation, coupling the
on-demand elimination to such a single-potential model is a natural next step.

\paragraph{Contributions} This work combines nonlinear elimination with a trust-region globalization and a
bound constraint, tailored to phase-field fracture:
\begin{itemize}
  \item a \emph{coupled crack-front nonlinear elimination}: the eliminated ``hard set'' spans
  \emph{both} fields (displacement and damage) and is selected \emph{algebraically} from the current
  residual and state, so no crack tracking is needed;
  \item a bound-constrained realization: the irreversibility constraint is carried by a single
  primal--dual active set applied consistently to the coupled operator \emph{and} the elimination
  sub-solves, and the outer step is globalized by a Steihaug--Toint trust region rather than a line
  search, which tolerates the indefinite softening Hessian;
  \item a \emph{merit-safeguarded} sweep that prevents the active set from oscillating during rapid
  propagation;
  \item an \emph{on-demand} activation of the elimination: the sweep runs only when the monolithic
  coupled step stalls --- a stalling reduced-residual contraction, or a rejected trust-region step ---
  from thresholds grounded on the monolithic runs' own convergence statistics, so its per-iteration
  surcharge is never paid where plain Newton already converges (a bounded-overhead default in our tests); and
  \item a \emph{subdomain-restricted assembly}: since the eliminated iterate differs only on the hard
  set, the elimination's finite-element assembly is restricted to the elements incident to it, so its
  cost scales with the hard set. The whole construction reads equivalently as a NEPIN specialization or
  as an MSPIN field sweep restricted to the hard set.
\end{itemize}
Section~\ref{sec:problem} states the problem; Section~\ref{sec:method} develops the method;
Section~\ref{sec:results} reports brittle and ductile mode-I benchmarks; Section~\ref{sec:discussion}
discusses the results and extensions. Implementation details are collected in \ref{app:impl}.

\section{Problem formulation}\label{sec:problem}

Per quasi-static load step, the total potential energy is minimized over the coupled DOF
vector $x=(u,d)$ --- displacement $u$ and phase field $d\in[0,1]$ --- subject to the step's Dirichlet
BCs and the \emph{irreversibility} bound (damage cannot heal):
\begin{equation}\label{eq:min}
  \min_{x}\ \Psi(x)
  \qquad\text{s.t.}\qquad
  \dold \;\le\; d \;\le\; 1 ,
\end{equation}
where $\dold$ is the converged phase field of the previous step (lower bound) and $1$ is the physical
upper bound. At an iterate $x$ define
\begin{align}
  \text{coupled residual (energy gradient)}\quad & \nabla\Psi(x)=R(x)=\begin{bmatrix}R_u\\ R_d\end{bmatrix},\\
  \text{coupled Hessian}\quad & A(x)=\nabla^2\Psi(x)=\dfrac{\partial R}{\partial x}
   =\begin{bmatrix}A_{00}&A_{01}\\ A_{10}&A_{11}\end{bmatrix}.
\end{align}
$A$ is \emph{symmetric} ($A_{01}=A_{10}\Tr$) and generally \emph{indefinite} near the limit point
(softening drives the coupled operator's eigenvalues negative even though the diagonal blocks
$A_{00},A_{11}$ stay symmetric positive definite, SPD). For the phase-field models considered here
$R=\nabla\Psi$ holds exactly, so the energy $\Psi$ is an exact merit; a residual norm
$\tfrac12\norm{R}^2$ is available as an alternative globalization (Section~\ref{sec:tr}). The bound in
\eqref{eq:min} makes each step a \emph{bound-constrained variational inequality}, not a smooth
root-find; the primal--dual active set of Section~\ref{sec:pdas} handles it.

\paragraph{The potential} Concretely, $\Psi$ is the total potential energy of a phase-field fracture
model --- a damage-degraded stored bulk energy plus the regularized fracture surface energy:
\begin{equation}\label{eq:psi-form}
  \Psi(x)=\int_\Omega\!\Big[\,
    \underbrace{g(d)\,\psi_e^{+}(\bm\varepsilon)+\psi_e^{-}(\bm\varepsilon)}_{\text{elastic (split)}}
    \;+\;\underbrace{g_p(d)\,\psi_p(\bar\varepsilon^{p})}_{\text{plastic (ductile only)}}
    \;+\;\underbrace{\dfrac{\mathcal{G}_c}{c_0\,\ell}\!\Big(\alpha(d)+\ell^2\,\lvert\nabla d\rvert^2\Big)}_{\text{fracture surface}}
  \,\Big]\,\mathrm dV ,
\end{equation}
with degradation function $g$, crack geometric function $\alpha$ and its normalization $c_0$, toughness
$\mathcal{G}_c$, and regularization length $\ell$; the stored plastic energy $\psi_p$ is present only for the
ductile (elasto-plastic) model. The specific constitutive choices --- the elastic tension/compression
split, the crack geometric and degradation functions, and the $J_2$ plasticity and hardening that supply
$\psi_p$ --- together with the strong form of the boundary-value problem are collected in
\ref{app:models}. The solver developed below uses \emph{only} the variational structure just
stated --- that $R=\nabla\Psi$ and that $A=\nabla^2\Psi$ has the $2\times2$ field-split block form; it is
otherwise agnostic to the constitutive model, which is why the same method applies unchanged across the
brittle and ductile testbeds.

\section{Method}\label{sec:method}

The solver is an inexact-Newton trust region on the coupled energy (Section~\ref{sec:tr}), whose step
is computed by a Steihaug--Toint truncated conjugate-gradient (CG) solve of the trust-region
subproblem (Section~\ref{sec:cg}) with an SPD field-split preconditioner (Section~\ref{sec:precond}).
Irreversibility is enforced by a primal--dual active set (Section~\ref{sec:pdas}). The nonlinear
preconditioner (Section~\ref{sec:nepin}) eliminates the crack-front hard set before each step. The
complete iteration is Algorithm~\ref{alg:outer} (Section~\ref{sec:full}).

\subsection{Outer trust-region inexact Newton}\label{sec:tr}

The outer loop globalizes a scalar merit $\phi$:
\begin{equation}
  \phi(x)=
  \begin{cases}
    \Psi(x), & \text{energy merit (\textbf{E}, default; $R=\nabla\Psi$)},\\[2pt]
    \tfrac12\,\norm{R(x)}^2, & \text{residual merit (\textbf{R}).}
  \end{cases}
\end{equation}
The globalization is a trust region; the merit (\textbf{E} or \textbf{R}) selects only the
\emph{acceptance} test, giving variants \textbf{TR-E} and \textbf{TR-R} that share one trust-region solve.
Both build the step from the \emph{same} Newton system $A\,p=-R$ (below): for \textbf{E} this is the exact
trust-region model of the energy, since $R=\nabla\Psi$ and $A=\nabla^2\Psi$; for \textbf{R} it is a
\emph{residual-merit-globalized}, Steihaug-truncated Newton --- the same Newton direction, then
globalized by the merit $\tfrac12\norm{R}^2$. It is \emph{not} the Gauss--Newton least-squares model
$\min_p\tfrac12\norm{R+Ap}^2$ (whose Hessian is $A\Tr A$ and gradient $A\Tr R$); we keep the coupled Newton
direction because the \emph{unconstrained} Newton step is a descent direction for $\tfrac12\norm{R}^2$ when
$A$ is symmetric and nonsingular --- $\nabla(\tfrac12\norm{R}^2)=A\Tr R=AR$ and
$(AR)\Tr(-A^{-1}R)=-\norm{R}^2<0$. The trust-region--truncated step need not preserve this: at the boundary
or under negative curvature the Steihaug step can fail to reduce $\tfrac12\norm{R}^2$, so the residual merit
is used only for acceptance and may reject otherwise useful energy-model steps in the indefinite regime ---
consistent with the fragility of \textbf{TR-R} in Section~\ref{sec:results}. Either way $\tfrac12\norm{R}^2$
enters only through the accept/reject ratio, the sole difference between the variants.

At iterate $x_k$ with radius $\Delta_k$, both variants build one quadratic model from the residual $R_k$
(the energy gradient $\nabla\Psi$, distinct from the degradation function $g(d)$) and the coupled Hessian
$A_k$,
\begin{equation}
  m_k(p)=\Psi(x_k)+R_k\Tr p+\tfrac12\,p\Tr A_k\,p,
\end{equation}
and take the step by approximately solving the trust-region subproblem~\citep{ConnGouldToint2000}
\begin{equation}\label{eq:trs}
  p_k=\arg\min_p\ m_k(p)\quad\text{s.t.}\quad \Pnorm{p}\le\Delta_k,
\end{equation}
where $\Pnorm{p}=(p\Tr P p)^{1/2}$ is induced by an SPD field-split preconditioner $P$
(Section~\ref{sec:precond}). Accept/adjust on the ratio of actual to predicted reduction,
\begin{equation}\label{eq:rho}
  \rho_k=\dfrac{\phi(x_k)-\phi(x_k+p_k)}{\pred_k},\qquad
  \pred_k=
  \begin{cases}
    -(R_k\Tr p_k+\tfrac12 p_k\Tr A_k p_k), & \text{energy merit},\\[4pt]
    \tfrac12\big(\norm{R_k}^2-\norm{R_k+A_k p_k}^2\big), & \text{residual merit.}
  \end{cases}
\end{equation}
With $\eta_1=0.1,\ \eta_2=0.75$, shrink factor $\tfrac14$, expand factor $2$: accept
$x_{k+1}=x_k+p_k$ if $\rho_k\ge\eta_1$; expand $\Delta$ if $\rho_k>\eta_2$ and the step is on the
boundary; otherwise shrink and re-solve \eqref{eq:trs}. Because the subproblem's optimality condition
$(A_k+\lambda P)p_k=-R_k$ with $\lambda\ge0$ and $A_k+\lambda P\succeq0$ remains well-posed for
indefinite $A_k$, the trust region handles the softening regime that a step along a single fixed
direction cannot.

\paragraph{Decoupled acceptance under bounds} With the irreversibility bound active, the energy
$\Psi$ is a poor ratio-test merit \emph{near} the constrained minimizer: $\Psi$ is essentially flat
there (already minimized over the inactive space) while the reduced KKT/complementarity residual
$\norm{R_\Iset}$ (Section~\ref{sec:pdas}) is not yet zero. Then $\rho_k$ (energy) degrades to noise and
rejects every step, collapsing $\Delta$. The two roles the merit was serving are \emph{decoupled}: the
energy $\Psi$ globalizes the smooth descent through $\rho_k$, while the reduced residual
$\norm{R_\Iset}$ measures the active-set/complementarity resolution. The step is accepted if
\emph{either} makes progress,
\begin{equation}\label{eq:accept}
  \rho_k\ge\eta_1 \quad\text{or}\quad \norm{R_\Iset(x_k+p_k)} < \norm{R_\Iset(x_k)},
\end{equation}
and $\Delta$ is shrunk only when both fail. The reduced-residual decrease is itself the safeguard
against an overshooting step, so neither merit vetoes the other, and this $\norm{R_\Iset}$ is the same
quantity a reduced-space variational-inequality solve monitors, so outer acceptance and inner
sub-solves use one consistent complementarity measure.

\paragraph{On the projected step and standard theory} Two departures from textbook trust-region theory
enter with the bound, and we state them plainly. (i)~The accepted trial is the \emph{projected} point
$W=\text{project}(x_k+p_k)$, whereas $\pred_k$ is the model reduction along the unprojected step $p_k$, so
$\rho_k$ is a faithful actual-over-predicted ratio only up to that projection. In practice the correction is
negligible: the active set is refreshed before each solve and changes by at most one DOF per outer iteration
(mean $0.64$, max $1$ over the AT1 history), and the reduced solve holds the active DOFs fixed, so $W$
differs from $x_k+p_k$ in at most that single boundary DOF. (ii)~The reduced-residual branch
of~\eqref{eq:accept} can admit an energy-flat or slightly energy-increasing step when the reduced KKT
residual $\norm{R_\Iset}$ strictly decreases. This is deliberate --- near the constrained minimizer $\Psi$
is flat while $\norm{R_\Iset}$ is not. A step accepted solely through this branch strictly decreases
$\norm{R_\Iset}$ relative to the current iterate --- a practical safeguard in the constrained regime --- but
this does \emph{not} imply a globally monotone reduced residual, since energy-accepted steps may raise
$\norm{R_\Iset}$. We do not claim a full convergence proof for the coupled
bound-constrained trust region: the standard Steihaug--Toint guarantees apply on the inactive space under
the energy ratio, and the reduced-residual branch is a feasibility-preserving safeguard for the degenerate
constrained regime; a complete analysis is left open.

\subsection{Trust-region subproblem: Steihaug--Toint truncated CG}\label{sec:cg}

The subproblem~\eqref{eq:trs} is solved approximately and matrix-free by preconditioned CG on the
Newton system $A\,p=-R$, truncated by the radius and by negative
curvature~\citep{Steihaug1983,Toint1981}. It needs only Hessian--vector products $v\mapsto Av$,
preconditioner solves $r\mapsto P^{-1}r$, and $P$-inner products (Algorithm~\ref{alg:steihaug}).

\begin{algorithm}[htbp]
\hrule\vspace{3pt}
\textbf{Algorithm~\refstepcounter{algorithm}\thealgorithm\label{alg:steihaug}: Steihaug--Toint preconditioned truncated CG for \eqref{eq:trs}}
\par\vspace{2pt}\hrule\vspace{3pt}
\footnotesize
\begin{tabbing}
\hspace*{1.8em}\=\hspace*{1.8em}\=\hspace*{2em}\=\kill
\kw{Input:} $g$; operators $A(\cdot),P^{-1}(\cdot)$; radius $\Delta$; forcing tol $\varepsilon$ \\
\kw{Output:} step $p$, predicted reduction $\pred$, boundary flag \\[2pt]
$p\gets 0$;\quad $r\gets g$;\quad $y\gets P^{-1}r$;\quad $d\gets -y$;\quad $\zeta\gets r\Tr y$ \quad($d$: CG direction) \\
\kw{if} $\norm{r}\le\varepsilon\norm{g}$ \kw{then return} $p=0$ \quad(already stationary)\\
\kw{for} $j=0,1,2,\dots$ \kw{do} \\
\> $\kappa\gets d\Tr A d$ \quad(curvature along $d$) \\
\> \kw{if} $\kappa\le 0$ \kw{then} \quad(a)~negative curvature $\to$ boundary \\
\>\> $\tau\gets$ positive root of $\Pnorm{p+\tau d}=\Delta$;\quad \kw{return} $p+\tau d$ \\
\> $\alpha\gets \zeta/\kappa$ \\
\> \kw{if} $\Pnorm{p+\alpha d}\ge\Delta$ \kw{then} \quad(b)~boundary hit \\
\>\> $\tau\gets$ positive root of $\Pnorm{p+\tau d}=\Delta$;\quad \kw{return} $p+\tau d$ \\
\> $p\gets p+\alpha d$;\quad $r\gets r+\alpha\,(A d)$ \\
\> \kw{if} $\norm{r}\le\varepsilon\norm{g}$ \kw{then return} $p$ \quad(c)~interior convergence \\
\> $y\gets P^{-1}r$;\quad $\zeta^{+}\gets r\Tr y$;\quad $\beta\gets \zeta^{+}/\zeta$;\quad $d\gets -y+\beta d$;\quad $\zeta\gets\zeta^{+}$ \\
Every \kw{return} above also reports $\pred=-(g\Tr p+\tfrac12 p\Tr A p)$ \quad(Steihaug guarantees $\pred>0$) \\
\end{tabbing}
\hrule
\end{algorithm}

\noindent The $j{=}0$ truncation is exactly the field-split direction $-P^{-1}g$; letting CG build the
Krylov subspace escapes a near-orthogonal single direction (the propagation regime). Forcing
$\varepsilon=\min(0.1,\sqrt{\norm{g}/\norm{g_0}})$ (Eisenstat--Walker~\cite{EisenstatWalker1996})
gives a superlinear tail. The boundary root is the positive solution of
$\Pnorm{p}^2+2\tau\langle p,d\rangle_P+\tau^2\Pnorm{d}^2=\Delta^2$.

\subsection{Field-split preconditioner}\label{sec:precond}

CG needs an SPD preconditioner and a symmetric operator. $A$ is symmetric. Writing $A=D+L+U$ with
$D=\mathrm{diag}(A_{00},A_{11})$, $L=\big[\begin{smallmatrix}0&0\\ A_{10}&0\end{smallmatrix}\big]$,
$U=L\Tr$, two consistent SPD choices are block-Jacobi $P=D$ (additive; $P^{-1}$ = the two inner block
solves in parallel) and symmetric block Gauss--Seidel; the latter is used,
\begin{equation}
  P_{\mathrm{SGS}}=(D+L)D^{-1}(D+U),
\end{equation}
which uses the coupling (a forward and backward block sweep) and is a stronger preconditioner, so
fewer CG iterations. The linear preconditioner is the SPD form of the \emph{same} splitting that
defines the nonlinear preconditioner: the multiplicative sweep $M=D+L$ has SPD symmetrization exactly
$P_{\mathrm{SGS}}=M D^{-1}M\Tr$, while the additive splitting $D$ pairs with block-Jacobi. Softening
keeps the diagonal blocks SPD (so $P$ is SPD and the inner solves are well-posed); the \emph{coupled}
indefiniteness is handled by Steihaug's negative-curvature exit, not by $P$.

\subsection{Irreversibility: primal--dual active set}\label{sec:pdas}

The bound $\dold\le d\le1$ is enforced by a primal--dual active set
(PDAS)~\citep{HintermullerItoKunisch2002,HeisterWheelerWick2015}: once per outer
iteration, the DOFs at which the bound is binding are identified and pinned, and the solve proceeds on the rest. The
identification is a projected-gradient (min-map) criterion~\citep{BensonMunson2006} --- current
position and multiplier sign, with no predictor --- made precise below and applied consistently at both
scales.

\paragraph{Optimality conditions} Since $R=\nabla\Psi$, the first-order (KKT) conditions of
$\min_d\Psi$ subject to $\dold\le d\le1$ are the componentwise complementarity system
\begin{equation}\label{eq:kkt}
  d_i=(\dold)_i\ \Rightarrow\ R_i\ge0,\qquad
  d_i=1\ \Rightarrow\ R_i\le0,\qquad
  (\dold)_i<d_i<1\ \Rightarrow\ R_i=0,
\end{equation}
equivalently the projected-gradient fixed point $d=P_{[\dold,1]}\!\big(d-R_d\big)$, where
$P_{[\dold,1]}$ is the componentwise projection onto the box and $R_d$ is the damage block of $R$. The
active set is the set of DOFs at which this projection is binding --- a DOF pinned at a bound with the
gradient pushing outward through it. The criterion below identifies that set at the current iterate.

\paragraph{Mesh-independent driving force} The finite-element residual
$R_i=\int_\Omega(\cdots)\,N_i\,\mathrm{d}V$ scales with the nodal control volume, so a fixed tolerance
on $|R_i|$ would be mesh-dependent. Let $B$ be the mass matrix on $d$, lumped by row sum
$B_{ii}=\int_\Omega N_i\,\mathrm{d}V$ (assembled once and cached); dividing by it gives the pointwise
\emph{driving force} $\lambda_i=(B^{-1})_{ii}R_i=R_i/B_{ii}$, whose units are mesh-independent. Because
$B_{ii}>0$, $\lambda_i$ and $R_i$ share a sign, so the scaling does not change \emph{which} DOFs are
identified --- only the mesh-independence of the tolerance $\lambda_{\mathrm{tol}}$ introduced next.
(Displacement DOFs are never bound-constrained.)

\paragraph{Active-set criterion} With a bound tolerance $\varepsilon_{\mathrm b}$ and a
\emph{dead-band} $\lambda_{\mathrm{tol}}\ge0$ (defined below), a $d$-DOF is classified from its position
and the sign of its driving force $\lambda_i$:
\begin{equation}\label{eq:pdas}
\begin{aligned}
  i\in\Aset^{\downarrow}\ (\text{lower})\ &\Longleftrightarrow\ d_i\le \dold+\varepsilon_{\mathrm b}\ \ \text{and}\ \ \lambda_i > +\lambda_{\mathrm{tol}} &&\Rightarrow\ \text{freeze } d_i=(\dold)_i,\\
  i\in\Aset^{\uparrow}\ (\text{upper})\ &\Longleftrightarrow\ d_i\ge 1-\varepsilon_{\mathrm b}\ \ \text{and}\ \ \lambda_i < -\lambda_{\mathrm{tol}} &&\Rightarrow\ \text{freeze } d_i=1,\\
  i\in\Iset\ (\text{inactive})\ &\Longleftrightarrow\ \text{otherwise,}
\end{aligned}
\end{equation}
and $\Aset=\Aset^{\downarrow}\cup\Aset^{\uparrow}$. With $\lambda_{\mathrm{tol}}=0$ this is exactly the
identification of the KKT active set~\eqref{eq:kkt}: a DOF is frozen iff it sits at a bound \emph{and}
its driving force points outward through it. There is \emph{no} predictor --- the test uses only the
current position and gradient, not a trial increment.

The \emph{dead-band} $\lambda_{\mathrm{tol}}$ is a tolerance interval
$[-\lambda_{\mathrm{tol}},+\lambda_{\mathrm{tol}}]$ around zero within which the driving force is treated
as numerically indifferent, so a DOF there is left inactive even when it sits at a bound. It is needed
because the sign-only test ($\lambda_{\mathrm{tol}}=0$) is ill-conditioned near incipient damage: a DOF
resting at $\dold$ with a driving force at the level of residual roundoff has an essentially arbitrary
sign, so it flickers in and out of $\Aset$ from one iteration to the next; the active set never settles
and the outer-iteration count inflates. Requiring $|\lambda_i|>\lambda_{\mathrm{tol}}$ before a DOF is
activated suppresses this oscillation, leaving the unstressed pre-damage region ($\lambda_i\approx0$)
inactive. Algorithm~\ref{alg:pdas} identifies $\Aset$ and projects onto the bounds.

\begin{algorithm}[htbp]
\hrule\vspace{3pt}
\textbf{Algorithm~\refstepcounter{algorithm}\thealgorithm\label{alg:pdas}: PDAS active-set update (identify $+$ project) at an iterate $x=(u,d)$}
\par\vspace{2pt}\hrule\vspace{3pt}
\footnotesize
\begin{tabbing}
\hspace*{1.8em}\=\hspace*{2em}\=\kill
\kw{Input:} $x=(u,d)$, residual $R=\nabla\Psi(x)$, lumped $B^{-1}$, lower bound $\dold$, dead-band $\lambda_{\mathrm{tol}}\ge0$, bound tol $\varepsilon_{\mathrm b}$ \\
\kw{Output:} active set $\Aset$; feasible $x$ (each active $d_i$ pinned to its bound) \\[2pt]
$\lambda\gets B^{-1}R_d$ \quad(driving force on the $d$-DOFs; $\lambda_i=R_i/B_{ii}$, same sign as $R_i$) \\
$\Aset^{\downarrow},\Aset^{\uparrow}\gets\varnothing$ \\
\kw{for} each $d$-DOF $i$ \kw{do} \\
\> \kw{if} $d_i\le(\dold)_i+\varepsilon_{\mathrm b}$ \kw{and} $\lambda_i>+\lambda_{\mathrm{tol}}$ \kw{then} $\Aset^{\downarrow}\!\gets\Aset^{\downarrow}\cup\{i\}$, \ $d_i\gets(\dold)_i$ \\
\> \kw{else if} $d_i\ge 1-\varepsilon_{\mathrm b}$ \kw{and} $\lambda_i<-\lambda_{\mathrm{tol}}$ \kw{then} $\Aset^{\uparrow}\!\gets\Aset^{\uparrow}\cup\{i\}$, \ $d_i\gets 1$ \\
\> \kw{else} $i\in\Iset$ \\
\kw{return} $\Aset=\Aset^{\downarrow}\cup\Aset^{\uparrow}$, \ feasible $x$ \\
\end{tabbing}
\hrule
\end{algorithm}

\paragraph{Reduced operator} Freezing $\delta d=0$ on $\Aset$ is a symmetric row/column
(dynamic-Dirichlet) elimination of $A$ and $R$: zero the active rows \emph{and} columns of $A_{11}$
with unit diagonal; zero the active rows of $A_{10}$ and the active columns of $A_{01}$; and zero the
active entries of $R_d$. Because the step's initial guess is $d=\dold$, freezing the \emph{increment}
holds active nodes at their feasible value. The convergence norm and $\pred/\ared$ use the
\emph{reduced} residual $R_\Iset$ (active entries zeroed).

\paragraph{Consistency at both scales} The constraint is enforced by eliminating the \emph{same}
active set $\Aset$ wherever damage is updated --- in the coupled operator (above) \emph{and} in the
field-split preconditioner's damage block. The damage sub-solve inside the nonlinear preconditioner
therefore holds $\delta d=0$ on $\Aset$ (each active node pinned at its feasible bound value) and
updates only the inactive DOFs $\Iset$, exactly as the coupled operator does; were it to move $d$ on
$\Aset$, it would drive damage across the bound and the two scales would disagree, defeating the
constraint. \emph{How} the pin is imposed is an implementation choice: the sub-solve is run as a
bound-constrained (variational-inequality) solve, but any sub-solve that eliminates the same active
DOFs as Dirichlet data --- e.g.\ a reduced-space solve on $\Iset$ --- yields the identical partition.

\paragraph{Convergence} Every trial step $W=x+p$ is projected onto the box
($d\mapsto\min(1,\max(\dold,d))$) before the merit is evaluated. Convergence is declared on the
reduced residual by two branches: \emph{absolute}, $\norm{R_{\Iset}}<\mathrm{atol}$, accepted
regardless of active-set motion (a reduced residual at the noise floor means any remaining set churn
involves borderline sub-tolerance DOFs); and \emph{relative}, $\Aset_{k}=\Aset_{k-1}$ and
$\norm{R_{\Iset}}<\mathrm{rtol}\,\norm{R_{\Iset,0}}$ --- a relative drop is a PDAS solution only once
the active set has settled~\citep[Remark~3.3]{HeisterWheelerWick2015}.

\subsection{Nonlinear-elimination preconditioner}\label{sec:nepin}

\paragraph{Nonlinear preconditioning} Each load step reduces to the coupled residual equation $R(x)=0$ for
$x=(u,d)$. In this problem the nonlinearity is strongly \emph{unbalanced} --- it localizes in the process zone at the
advancing crack front while the far field responds nearly linearly --- and a plain coupled Newton then
develops the long residual plateau documented in Section~\ref{sec:intro}.
Nonlinear preconditioning~\citep{BruneKnepleySmithTu2015} addresses this by replacing $R(x)=0$ with an
equivalent problem that Newton finds easier, built from an inexpensive nonlinear map $\mathrm{NPC}$ that
leaves the solution invariant --- $R(x^\star)=0\Rightarrow\mathrm{NPC}(x^\star)=x^\star$, and for a sweep over
the \emph{whole} system the converse holds as well, so its fixed points are exactly the roots of $R$. There are
two ways to use such a map. In the \emph{left} (preconditioned-residual) form of ASPIN and
MSPIN~\citep{CaiKeyes2002,LiuKeyes2015}, Newton is applied not to $R$ but to the fixed-point residual
\begin{equation}\label{eq:fspin}
  F(x)\;=\;x-\mathrm{NPC}(x),
\end{equation}
which has the same root $x^\star$ but a far more balanced nonlinearity, because $\mathrm{NPC}$ has already
resolved the strong local part; each correction solves $F'(x)\,\delta=-F(x)$ with Jacobian
$F'(x)=I-\mathrm{NPC}'(x)$, which has the familiar local linearization $F'(x)\approx M^{-1}A$ ($A=R'(x)$
the true coupled Jacobian and $M$ the sweep's block operator, both given below). In the \emph{right}
(nonlinear-elimination) form that we adopt~\citep{LanzkronRoseWilkes1996,CaiLi2011,LiuEtAl2022}, the map is
instead used to produce an improved iterate $\tilde x=\mathrm{NPC}(x)$, from which the ordinary step on the
\emph{original} $R$ is taken. The two are not interchangeable here. The left form redefines the residual ---
$F$ in \eqref{eq:fspin} is neither symmetric nor the gradient of any potential --- so it would forfeit the
variational structure the energy trust region depends on (Section~\ref{sec:intro}); the right form leaves
$R=\nabla\Psi$ untouched, so the Steihaug--Toint globalization of Section~\ref{sec:tr} carries over verbatim
and the preconditioner enters only through a better starting iterate. The rest of this section constructs
$\mathrm{NPC}$ as a bound-constrained field-split sweep, restricts it to the hard set, and takes the outer
step from $\tilde x$.

\paragraph{Field-split sweep} Concretely, $\mathrm{NPC}$ is a multiplicative block Gauss--Seidel sweep over
the two fields~\citep{LiuKeyes2015}: solve the $u$-block given the current $d$, then the $d$-block given the
updated $u$. Its fixed-point residual $F$ has the local linearization $F'(x)\approx M^{-1}A$, while the
sweep itself has local error-propagation operator $\mathrm{NPC}'(x)\approx I-M^{-1}A$, with
block-lower-triangular $M=D+L$ ($D$ the diagonal field blocks, $L$ the lower coupling block); $M$ is
nonsymmetric, consistent with $F$ in \eqref{eq:fspin} but disqualifying $M^{-1}$ as the symmetric-CG
preconditioner of Section~\ref{sec:precond}. Setting $\mathrm{NPC}=\mathrm{Id}$ recovers the monolithic (MONO) family;
sweeping the whole field is MSPIN. To keep the terminology straight: this \emph{nonlinear} field-split
sweep (the map $\mathrm{NPC}$) is distinct from the \emph{linear} field-split preconditioner $P$ of
Section~\ref{sec:precond} that accelerates the inner Steihaug CG; restricting the nonlinear sweep to the
hard set (below) is what we call NEPIN.

\paragraph{Hard set} The premise of nonlinear elimination is that the nonlinearity concentrates in a
small ``hard'' set of DOFs (physically, the process zone around the advancing crack) while the rest
responds nearly linearly; preconditioning only that set recovers most of the benefit at a fraction of
the work. The DOFs are partitioned into a ``hard'' (strongly nonlinear) set $\mathcal H$ and an ``easy''
(near-linear, far-field) set $\mathcal E$, chosen \emph{algebraically} per field on the current
iterate:
\begin{equation}\label{eq:hardset}
  \mathcal H_d = \{\, i : d_{\mathrm{lo}} < d_i < 1-d_{\mathrm{hi}} \,\}, \qquad
  \mathcal H_u = \{\, i : |R_{u,i}| > \tau \max_j |R_{u,j}| \,\},
\end{equation}
i.e.\ the damage process-zone band and the displacement force-imbalance front (thresholds
$d_{\mathrm{lo}},d_{\mathrm{hi}},\tau$); $\mathcal H=\mathcal H_u\cup\mathcal H_d$. Because the hard set is identified from the
current residual and state, the scheme needs no crack tracking and degrades gracefully when the hard
region is in fact broad.

\paragraph{Approximate elimination} Freezing $x_{\mathcal E}$, the \emph{ideal} elimination would solve the
hard set to convergence,
\begin{equation}\label{eq:elim}
  \text{find } x_{\mathcal H} \ \text{s.t.}\ R_{\mathcal H}(x_{\mathcal H}, x_{\mathcal E})=0 ,
\end{equation}
returning $x_{\mathcal H}$ with $x_{\mathcal E}$ held fixed. We do not solve~\eqref{eq:elim} exactly: one
application of $\mathrm{NPC}$ is a \emph{single} multiplicative field-split sweep over the hard set --- one
bound-constrained displacement-front solve given the current $d$, then one damage-band solve given the
updated $u$ --- so it is an \emph{approximate} nonlinear elimination, equivalently a restricted nonlinear
block relaxation~\citep{CaiLi2011,LiuEtAl2022}. Because the damage update generally perturbs the
displacement residual, $R_{\mathcal H}$ is not driven to zero, and a fixed point of the restricted map need
not be a root of the full $R$. This is immaterial to correctness: the outer Steihaug--Toint step is always
taken on the \emph{original} $R$ (Section~\ref{sec:tr}), and the sweep supplies only a better starting
iterate $\tilde x=\mathrm{NPC}(x)$. Each block sub-solve is a reduced-space, bound-constrained Newton solve
with the far field held as Dirichlet data and $d\ge\dold$ carried inside the damage sub-solve exactly as in
Section~\ref{sec:pdas}. Coupling between the fields is transmitted through the sequential material updates
(the displacement solve sees the current $d$; the damage solve sees the updated $u$), so the off-diagonal
blocks are \emph{not} used in the sweep --- they appear only in the outer operator $A$.

\paragraph{Variants} Choosing what the sweep eliminates recovers a family:
$\mathrm{NPC}=\mathrm{Id}$ is \textbf{MONO}; sweeping the whole field is \textbf{MSPIN}; eliminating the
coupled hard set $\mathcal H_u\cup\mathcal H_d$ (both fields) is \textbf{NEPIN}. Eliminating the
two fields \emph{together} is essential --- eliminating damage alone would drive $d$ ahead of a frozen
$u$ into a coupled-inconsistent state the outer trust region must undo. Two equivalent readings of
NEPIN: the restricted sweep is a NEPIN specialization (approximate nonlinear elimination of a hard set,
here spanning two coupled fields under a bound constraint), or an MSPIN extension (the multiplicative
field-split sweep restricted to the hard set).

\paragraph{On-demand activation} The sweep is not free: each application adds the block sub-solves ---
and their residual/Jacobian assemblies --- on top of the monolithic step. Applying it every outer
iteration therefore taxes the iterations that do not need it --- the elastic and post-crack steps, where
plain coupled Newton already contracts quadratically --- while only the process-zone iterations repay the
surcharge (Section~\ref{sec:discussion} quantifies this in assembly counts). The sweep is thus activated
\emph{on demand}, from the monolithic step's own convergence. With the reduced-residual contraction
$\theta_k=\norm{R_{\Iset}(x_k)}/\norm{R_{\Iset}(x_{k-1})}$, the elimination is switched \emph{on} when the
coupled Newton stalls --- $\theta_k>\theta_{\mathrm{on}}$ for two consecutive (post-load-increment)
iterations, or the trust region rejects a trial (its quadratic model has failed) --- and \emph{off} once
contraction resumes, $\theta_k<\theta_{\mathrm{off}}$; the hysteresis gap
$\theta_{\mathrm{off}}<\theta_{\mathrm{on}}$ prevents chatter, and the first post-increment ratio (a
transient jump, not a stall) is skipped. The thresholds are not tuned per problem: on monolithic runs the
contraction cleanly separates converging iterations ($\theta$ near $0$) from stalling ones
($\theta\!\gtrsim\!0.8$ at nucleation), and rejections fall exclusively on the hard steps --- fixing
$\theta_{\mathrm{on}}\approx0.5$, $\theta_{\mathrm{off}}\approx0.1$ across both the brittle and ductile
testbeds. On-demand activation makes the elimination a \emph{bounded-overhead} default in our experiments: it collapses to
monolithic Newton (no surcharge) wherever the coupled step is healthy and engages only in the
crack-front regime it was designed for.

\paragraph{Merit-safeguarded sweep} The sweep with active set $\Aset$ is a semismooth-Newton block
step. Under rapid front propagation the set $\Aset$ (identified at the pre-sweep iterate) can freeze
DOFs that should advance, so $\mathrm{NPC}(x)$ may \emph{raise} the KKT residual; adopting it
unconditionally --- it sits outside the trust region's ratio test~\eqref{eq:rho} --- lets the active
set oscillate and the outer iteration settle into a limit cycle. The sweep is globalized by a
backtracking line search on the KKT-residual merit
\begin{equation}\label{eq:kktmerit}
  \varphi(x)=\tfrac12\,\norm{R_{\Iset}(x)}^2 ,
\end{equation}
deliberately \emph{not} the energy $\Psi$: a block minimization can lower $\Psi$ while raising
$\norm{R}$, so $\Psi$ is blind to the oscillation, whereas $\varphi$ measures exactly the
stationarity/complementarity residual it degrades. With sweep direction $s=\mathrm{NPC}(x)-x$, accept
the largest $t\in\{1,\tfrac12,\dots\}$ with $\varphi(x+ts)\le\varphi(x)$; if none (a pure-ascent
sweep), take $t=0$, so the ratio-guarded trust-region step makes the progress instead
(Algorithm~\ref{alg:sweep}). This prevents the NPC application itself from amplifying the KKT-residual
oscillation; the subsequent outer trust-region step remains governed by the acceptance
rule~\eqref{eq:accept}. This safeguard applies to the elimination families only; MONO takes no sweep.

\begin{algorithm}[htbp]
\hrule\vspace{3pt}
\textbf{Algorithm~\refstepcounter{algorithm}\thealgorithm\label{alg:sweep}: Merit-safeguarded sweep at iterate $x$ with active set $\Aset$ (inactive $\Iset$)}
\par\vspace{2pt}\hrule\vspace{3pt}
\footnotesize
\begin{tabbing}
\hspace*{1.8em}\=\hspace*{2em}\=\kill
\kw{Input:} $x$, active set $\Aset$, KKT merit $\varphi(\cdot)=\tfrac12\norm{R_{\Iset}(\cdot)}^2$, max backtracks $b_{\max}$ \\
\kw{Output:} globalized iterate $x$ \\[2pt]
$\varphi_0\gets\varphi(x)$;\quad $x_{\mathrm{pre}}\gets x$ \\
$x\gets\mathrm{NPC}(x)$ with $\Aset$ pinned in the damage sub-solve;\quad $s\gets x-x_{\mathrm{pre}}$;\quad $t\gets1$;\quad $b\gets0$ \\
\kw{while} $\varphi(x)>\varphi_0$ \kw{and} $b<b_{\max}$ \kw{do} \quad $t\gets t/2$;\quad $x\gets x_{\mathrm{pre}}+t\,s$;\quad $b\gets b+1$ \\
\kw{if} $\varphi(x)>\varphi_0$ \kw{then} $x\gets x_{\mathrm{pre}}$ \quad(reject pure-ascent sweep: $t=0$) \\
\kw{return} $x$ \\
\end{tabbing}
\hrule
\end{algorithm}

\paragraph{Subdomain-restricted assembly} Because $\tilde x$ differs from $x$ only on $\mathcal H$,
the reduced-space linear solve already touches only the free DOFs; the remaining full-domain cost is
the finite-element \emph{assembly} of $R$ and $A$. It is restricted to the elements incident to
$\mathcal H$: the free rows are then fully assembled while the far field is untouched, so the
elimination's assembly scales with $\mathcal H$. The restriction changes only cost, not the iterates
--- they are identical to full-domain assembly to numerical precision. The magnitude of the saving depends on how
localized $\mathcal H$ is.

\subsection{Full algorithm}\label{sec:full}

\begin{figure}[htbp]
\centering
\begin{tikzpicture}[
  font=\footnotesize,
  >={Latex[length=2mm]},
  proc/.style={rectangle, rounded corners=2pt, draw, align=left, inner sep=4pt, text width=66mm},
  dec/.style={diamond, aspect=2.4, draw, align=center, inner sep=1pt},
  term/.style={rectangle, rounded corners=9pt, draw, align=center, inner sep=5pt, fill=black!4},
  lbl/.style={font=\scriptsize\itshape, text=black!55},
]
\node[term, text width=66mm] (setup) {Initialize $x_0$ (BC-satisfying), $\dold$;\ \ lumped mass $B^{-1}$ and radius $\Delta$ once per run};
\node[proc, below=7mm of setup] (res) {$R \gets \nabla\Psi(x)$};
\node[proc, below=5mm of res] (pdas) {PDAS update --- identify $+$ pin active set $\Aset$ \hfill{\footnotesize\textcolor{black!55}{(Alg.~\ref{alg:pdas})}}};
\node[dec, below=6mm of pdas] (conv) {$\norm{R_\Iset}<\text{tol}$?};
\node[proc, below=7mm of conv] (sweep) {Merit-safeguarded field-split sweep \hfill{\footnotesize\textcolor{black!55}{(Alg.~\ref{alg:sweep})}}\\[1pt] {\scriptsize MONO: skip \ \textbullet\ MSPIN: whole field \ \textbullet\ NEPIN: hard set $\mathcal H$}};
\node[proc, below=5mm of sweep] (asm) {Assemble coupled $A(x)$;\ eliminate active set $\Aset$};
\node[proc, below=9mm of asm] (steihaug) {$p \gets$ Steihaug--Toint truncated CG \hfill{\footnotesize\textcolor{black!55}{(Alg.~\ref{alg:steihaug})}}};
\node[proc, below=5mm of steihaug] (accept) {$W\gets\text{project}(x+p)$;\ ratio $\rho$ / KKT test $\Rightarrow$ accept \& grow, else shrink $\Delta$};
\node[term, right=34mm of conv, text width=24mm] (done) {converged $\Rightarrow$ next load step};

\draw[->] (setup)--(res);
\draw[->] (res)--(pdas);
\draw[->] (pdas)--(conv);
\draw[->] (conv)--node[lbl,left]{no}(sweep);
\draw[->] (conv)--node[lbl,above]{yes}(done);
\draw[->] (sweep)--(asm);
\draw[->] (asm)--(steihaug);
\draw[->] (steihaug)--(accept);
\draw[->] (accept.east) -- ++(10mm,0) |- node[lbl,pos=0.25,right,fill=white,inner sep=1pt]{reject} (steihaug.east);
\draw[->] (accept.west) -- ++(-16mm,0) |- node[lbl,pos=0.4,left,fill=white,inner sep=1pt]{accept: $x\gets W$} (res.west);

\begin{scope}[on background layer]
\node[draw, dashed, rounded corners, fit=(steihaug)(accept), inner sep=6pt] (trsbox) {};
\node[lbl, anchor=south west] at (trsbox.north west) {trust-region subproblem (accept/reject loop)};
\node[draw, rounded corners, fit=(res)(pdas)(conv)(sweep)(asm)(trsbox), inner sep=11pt] (outerbox) {};
\node[lbl, anchor=south west] at (outerbox.north west) {outer inexact-Newton iteration $k$};
\end{scope}
\end{tikzpicture}
\caption{Control flow of the full algorithm for one load step (Algorithm~\ref{alg:outer}), composing the
PDAS active-set update (Alg.~\ref{alg:pdas}), the merit-safeguarded field-split sweep
(Alg.~\ref{alg:sweep}; the family selects whether and over which set it runs), coupled-operator assembly
with active-set elimination, and the Steihaug--Toint trust-region subproblem (Alg.~\ref{alg:steihaug})
under the decoupled acceptance test~\eqref{eq:accept}. MONO skips the sweep; MSPIN sweeps the whole
field; NEPIN restricts it to the hard set $\mathcal H$.}
\label{fig:algmap}
\end{figure}
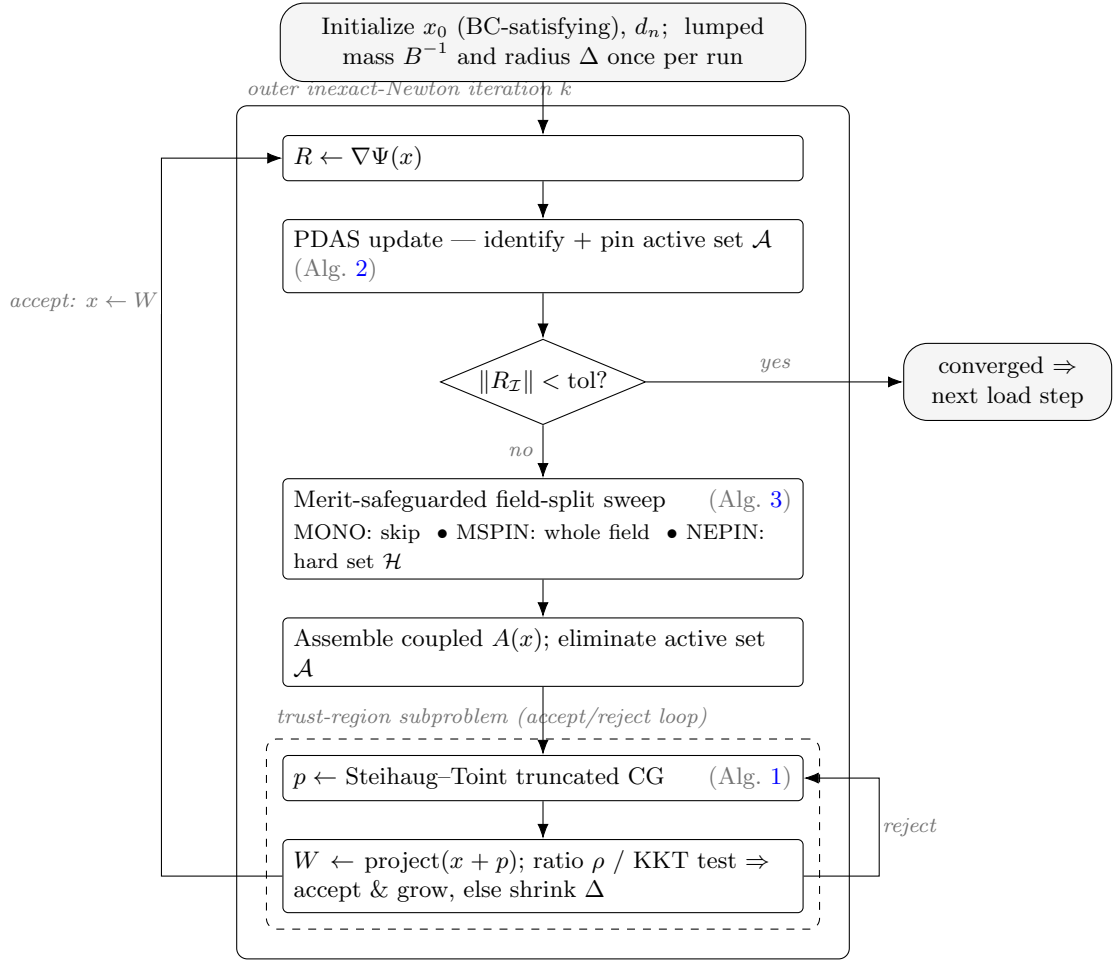

Algorithm~\ref{alg:outer} assembles the pieces of this section into the outer solve for a single load
step. Each iteration refreshes the residual $R=\nabla\Psi$ and the active set by the PDAS update
(Alg.~\ref{alg:pdas}); tests convergence on the reduced residual $\norm{R_\Iset}$ (with the
active-set-stationarity clause of Section~\ref{sec:pdas}); for the MSPIN and NEPIN families applies one
merit-safeguarded field-split sweep (Alg.~\ref{alg:sweep}), restricted to the hard set $\mathcal H$ for
NEPIN; then eliminates the active set from the coupled operator and takes a Steihaug--Toint
trust-region step under the decoupled acceptance test~\eqref{eq:accept}. The lumped mass $B^{-1}$ and
the trust radius are initialized once per solve; the family (MONO / MSPIN / NEPIN) selects only
whether, and over which set, the sweep runs. Figure~\ref{fig:algmap} shows the resulting control flow
schematically.

\begin{algorithm}[htbp]
\hrule\vspace{3pt}
\textbf{Algorithm~\refstepcounter{algorithm}\thealgorithm\label{alg:outer}: outer solve for one load step}
\par\vspace{2pt}\hrule\vspace{3pt}
\footnotesize
\begin{tabbing}
\hspace*{1.2em}\=\hspace*{1.2em}\=\hspace*{1.2em}\=\kill
\kw{Input:} $x_0$ (BC-satisfying), $\dold$, $(\mathrm{atol},\mathrm{rtol})$, dead-band $\lambda_{\mathrm{tol}}$, merit $\phi\in\{$E,R$\}$, \\
\> family (MONO / MSPIN / NEPIN), on-demand gate on/off with $(\theta_{\mathrm{on}},\theta_{\mathrm{off}})$ \\
Assemble/cache lumped mass $B^{-1}$ on $d$ (once per run);\quad $\Aset_{-1}\gets\varnothing$;\quad $\Delta\gets -1$ (uninitialized) \\
\kw{for} $k=0,1,2,\dots$ \kw{do} \\
\> $R\gets\nabla\Psi(x_k)$ \\
\> $(\Aset_k,\,x_k)\gets$ \textsc{pdas-update}$(x_k,\,R,\,\lambda_{\mathrm{tol}})$ \quad(Alg.~\ref{alg:pdas}) \\
\> $R_{\Iset}\gets R$ restricted to $\Iset_k$;\quad on $k{=}0$ record $\norm{R_{\Iset,0}}$ \\
\> \kw{if} $\norm{R_{\Iset}}<\mathrm{atol}$ \ \kw{or}\ $\big(\Aset_k=\Aset_{k-1}$ \kw{and} $\norm{R_{\Iset}}<\mathrm{rtol}\,\norm{R_{\Iset,0}}\big)$ \kw{then return} converged \\
\> \emph{on-demand gate} (if enabled): $\theta\gets\norm{R_{\Iset}}/\norm{R_{\Iset}^{\text{prev}}}$; \\
\>\> set \textsc{sweep-on} if $\theta>\theta_{\mathrm{on}}$ (twice) or last iter's TR rejected, clear if $\theta<\theta_{\mathrm{off}}$ \\
\> \kw{if} family uses a sweep \kw{and} (gate disabled \kw{or} \textsc{sweep-on}) \kw{then} \\
\>\> $x_k\gets$ \textsc{safeguarded-sweep}$(x_k,\Aset_k)$ (Alg.~\ref{alg:sweep}, restricted to $\mathcal H$ for NEPIN); \ $R\gets\nabla\Psi(x_k)$ \\
\> assemble $A(x_k)$; \ eliminate $\Aset_k$ from $(A,R)$ \quad(reduced operator, Section~\ref{sec:pdas}) \\
\> $\phi_k\gets\phi(x_k)$ \\
\> \kw{if} $\Delta\le 0$ \kw{then} $\Delta\gets\Pnorm{P^{-1}R}$ \\
\> \kw{repeat} \\
\>\> $p\gets$ Steihaug--TRS$(R,A,P^{-1},\Delta,\varepsilon)$ (Alg.~\ref{alg:steihaug}) \\
\>\> $W\gets$ project$(x_k+p)$;\ \ $\rho\gets(\phi_k-\phi(W))/\pred$ \\
\>\> \emph{accept} \kw{if} $\rho\ge\eta_1$ \kw{or} $\norm{R_{\Iset}(W)}<\norm{R_{\Iset}(x_k)}$ \ (grow $\Delta$ if $\rho>\eta_2$); \ \kw{else} shrink $\Delta$ \ \eqref{eq:accept} \\
\> \kw{until} accepted \kw{or} radius floor \\
\> $x_{k+1}\gets W$;\quad $\Aset_{k-1}\gets\Aset_k$ \\
\end{tabbing}
\hrule
\end{algorithm}

On persistent rejection the radius collapses; if $\Delta<\Delta_{\min}$ the step fails and the outer
time-stepping driver cuts $\Delta t$. Unlike a residual-norm merit, a trust-region rejection means the
model was wrong \emph{at this radius}, and shrinking recovers a good step --- so cutbacks are rare
rather than a spiral.

\section{Results}\label{sec:results}

\paragraph{Testbed} The primary comparison is between monolithic Newton (MONO) and the coupled-front
nonlinear elimination (NEPIN), both under the energy merit (TR-E), on one shared bound-constrained
mode-I problem in three settings: two brittle regularizations (AT2, AT1) and a strongly-coupled ductile
(elasto-plastic with plastic-work-driven damage --- the ``E-P-PD'' model, Table~\ref{tab:physics},
\ref{app:models}). The full-field
multiplicative sweep (MSPIN), of which NEPIN is the hard-set restriction, is compared separately in
Section~\ref{sec:res-mspin}.
The outer solve converges on the reduced coupled residual
$\norm{R_\Iset}$. The physical setup is described next; the discretization and
material parameters are collected in Table~\ref{tab:physics} and the algorithmic parameters in
Table~\ref{tab:params}.

\paragraph{Problem setup} The specimen is a single-edge-notched tension (SENT) square under mode-I
loading, modeled on its upper half by symmetry about the crack plane $y=0$: the domain is the rectangle
$[0,1]\times[0,0.5]$ (consistent units). The crack plane carries the initial notch as a traction-free
segment $\{y=0,\ 0\le x\le\tfrac12\}$ (an edge crack of length $a=\tfrac12$, half the width), while the
remaining ligament $\{y=0,\ \tfrac12\le x\le1\}$ enforces the symmetry condition $u_y=0$. The top edge
$y=0.5$ is pulled in tension by a prescribed vertical displacement $u_y=t$ (with $u_x=0$), ramped
monotonically over $120$ equal steps of $\Delta t=5\times10^{-5}$ to $u_y=6\times10^{-3}$; a fail-fast
floor $\Delta t_{\min}=10^{-6}$ makes a stalling solver reveal itself as a death-spiral rather than
masking it with cutbacks. The bulk is linear isotropic elasticity with a spectral tension/compression
split, so only the tensile elastic energy is degraded. Fracture is run in both standard regularizations
--- AT2 ($\alpha=d^2$, $c_0=2$) and AT1 ($\alpha=d$, $c_0=\tfrac{8}{3}$) --- each with a quadratic
degradation $g(d)=(1-d)^2(1-\eta)+\eta$. The two probe complementary regimes: AT2 has no elastic
threshold, so its damage field is diffuse and stresses the hard-set selection and the assembly
restriction, whereas AT1's elastic threshold localizes damage naturally into a compact band.
Irreversibility $\dold\le d\le1$ is imposed by the PDAS of Section~\ref{sec:pdas} with \emph{no} penalty
or viscous term, so $R=\nabla\Psi$ holds exactly and the energy merit is well-defined. The mesh is fixed
and conforming --- no adaptive refinement, so the
off-diagonal field coupling stays free of hanging-node constraints --- and deliberately coarse
(about $2.4$ elements across the regularization length): it under-resolves the crack band and is
intended for \emph{relative} solver comparison, not quantitative fracture.

\paragraph{Deliberately hard testbed} Two standard engineering remedies would ease the softening
regime; both are \emph{disabled}, so the benchmark discriminates the solvers rather than the remedies.
(i)~A small \emph{viscous} (rate-dependent) term in the damage evolution smooths the softening
instability and selects a solution branch at the bifurcation, better-conditioning the otherwise
indefinite Hessian; the viscosity is set to zero, keeping the step rate-independent. (ii)~A staggered
solve can be \emph{capped} at a fixed number of alternate-minimization sweeps and its result accepted
even if the coupled residual has not converged --- each subproblem is individually solved, so the
remaining error is only the inter-field coupling, which is often acceptable in practice; the full coupled
residual $\norm{R_\Iset}$ is instead required to meet the tolerance at every step, with no sweep
cap. Both remedies are legitimate in production, but here they would mask the differences between the
methods, so they are omitted to keep the softening bifurcation genuinely hard.

\begin{figure}[htbp]
\centering
\includegraphics[width=0.49\textwidth]{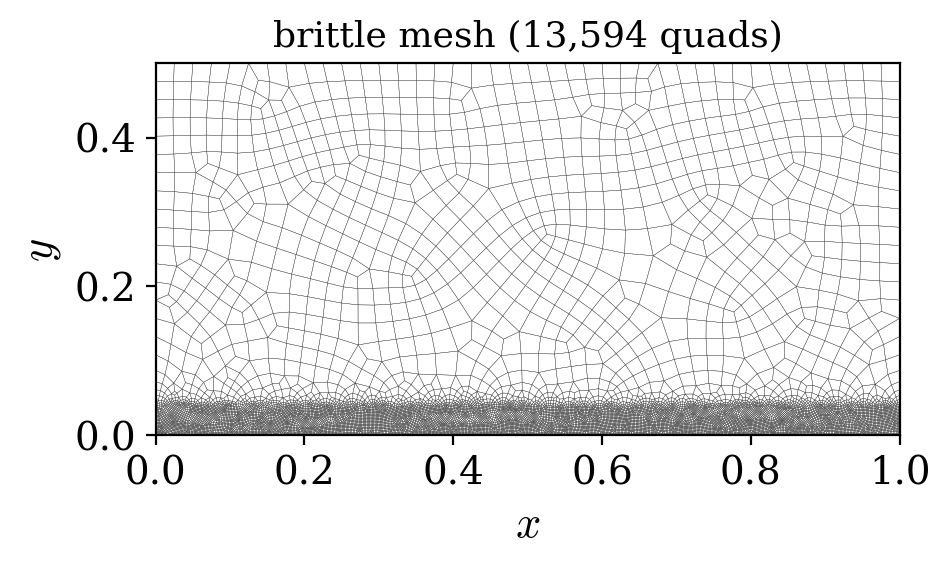}\hfill
\includegraphics[width=0.49\textwidth]{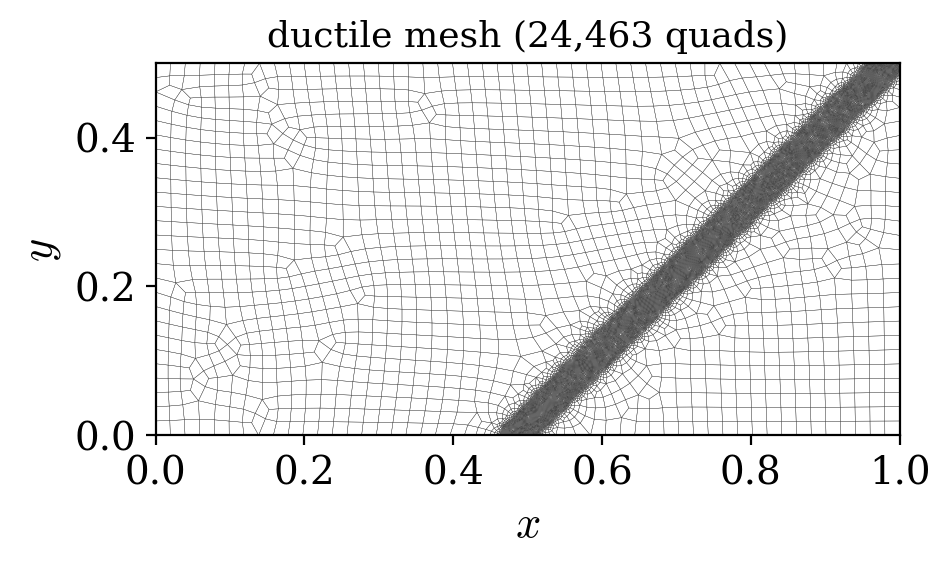}
\caption{Statically-refined conforming meshes, adapted to each testbed's crack path: (left) brittle ---
a horizontal band along the straight mode-I crack ($y=0$); (right) ductile --- a diagonal band along the
$\sim\!45^\circ$ slant crack from the notch tip. No adaptive refinement is used.}\label{fig:meshes}
\end{figure}

\begin{figure}[htbp]
\centering
\includegraphics[width=0.49\textwidth]{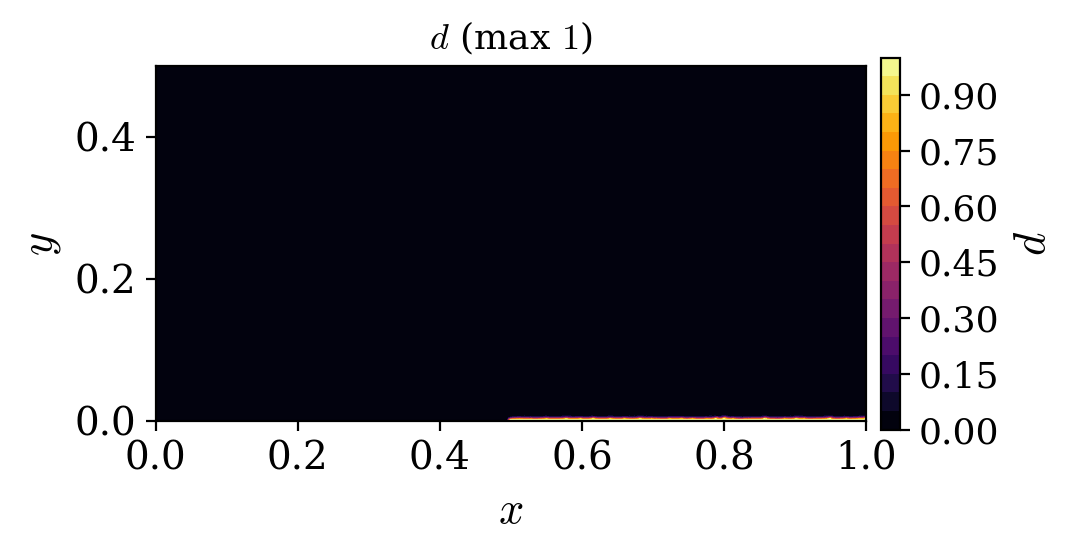}\hfill
\includegraphics[width=0.49\textwidth]{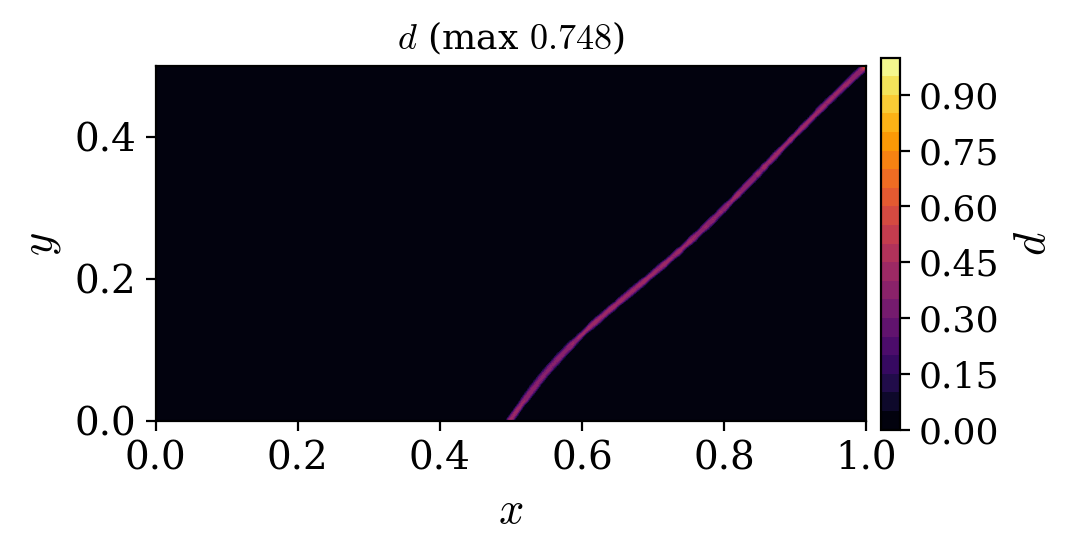}
\caption{Phase field $d$ at the final step: (left) brittle --- a straight mode-I crack along $y=0$;
(right) ductile --- a $\sim\!45^\circ$ slant crack following the plastic slip band. Same geometry and
regularization (AT1); only the constitutive model differs.}\label{fig:crack}
\end{figure}

\begin{figure}[htbp]
\centering
\includegraphics[width=0.49\textwidth]{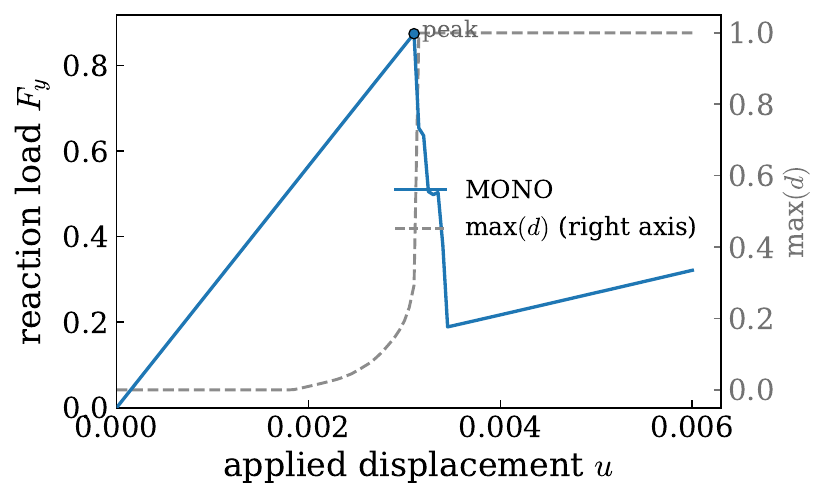}\hfill
\includegraphics[width=0.49\textwidth]{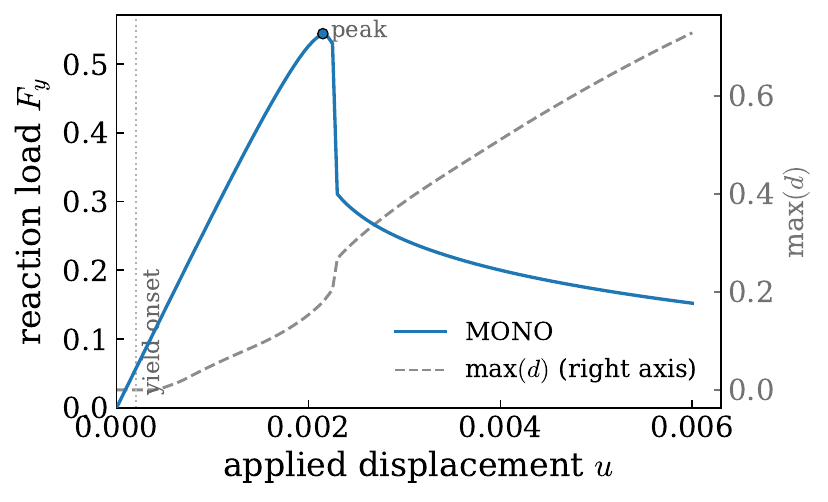}
\caption{Load--displacement response ($F_y$ vs.\ $u$): (left) brittle --- elastic rise to a peak then a
sharp drop as the crack forms; (right) ductile --- elastic, yield, plastic hardening, peak, post-peak
softening. Max damage overlaid on the right axis; identical for MONO and NEPIN.}\label{fig:load-disp}
\end{figure}

\begin{figure}[htbp]
\centering
\includegraphics[width=0.7\textwidth]{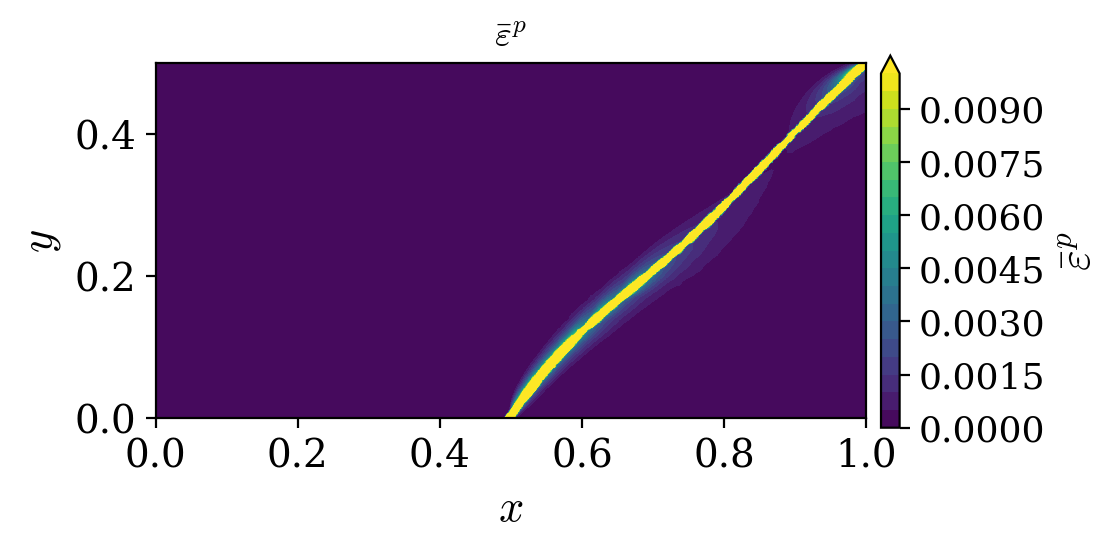}
\caption{Ductile testbed: effective plastic strain $\bar\varepsilon^{p}$ at the final step (color capped
at $0.01$ to reveal the zone). Plastic flow localizes into the
$\sim\!45^\circ$ band but spreads more broadly than the damage band --- the damage localization needs the
fine mesh, the diffuse plastic zone does not.}\label{fig:ep}
\end{figure}

\paragraph{The two testbeds} Beyond the two brittle regularizations, the same geometry and loading are
run with the strongly-coupled ductile (E-P-PD) model (\ref{app:models}). The problems differ
qualitatively: the brittle case fails by a straight mode-I crack along $y=0$, whereas the ductile case
--- whose plastic work drives damage --- fails by a $\sim\!45^\circ$ slant crack along the plastic slip
band from the notch tip (Figures~\ref{fig:crack},~\ref{fig:ep}; the upper-half symmetry model resolves
one arm of the symmetric slip-band pair), with a ductile hardening-then-softening
load response rather than a brittle snap (Figure~\ref{fig:load-disp}). Each mesh is statically refined
along its own crack path (Figure~\ref{fig:meshes}).

\begin{table}[htbp]
\centering
\begin{tabular}{lll}
\toprule
quantity & symbol & value \\
\midrule
\multicolumn{3}{l}{\emph{Geometry and loading (upper-half model, consistent units)}}\\
domain                       & ---              & $[0,1]\times[0,0.5]$ \\
initial crack length         & $a$              & $0.5$ (half width) \\
prescribed top displacement  & $u_y$            & $0\to6\times10^{-3}$ \\
load steps / step size       & ---              & $120$ / $\Delta t=5\times10^{-5}$ \\
fail-fast floor              & $\Delta t_{\min}$& $10^{-6}$ \\
\midrule
\multicolumn{3}{l}{\emph{Material (linear isotropic, spectral split)}}\\
Lam\'e parameters            & $\lambda,\mu$    & $121.15,\ 80.77$ \\
equivalently                 & $E,\nu$          & $210,\ 0.3$ \\
\midrule
\multicolumn{3}{l}{\emph{Fracture (two regularizations: AT2 and AT1)}}\\
crack geom.\ function, AT2    & $\alpha,\ c_0$   & $d^2,\ 2$ \\
crack geom.\ function, AT1    & $\alpha,\ c_0$   & $d,\ \tfrac{8}{3}$ \\
critical energy release rate & $\mathcal{G}_c$            & $2.7\times10^{-3}$ \\
regularization length        & $\ell$           & $3\times10^{-3}$ \\
residual stiffness           & $\eta$           & $10^{-3}$ \\
\midrule
\multicolumn{3}{l}{\emph{Discretization, brittle testbed (fixed conforming quadrilaterals)}}\\
elements / nodes             & ---              & $13{,}594$ / $13{,}909$ \\
DOFs (displacement / damage) & ---              & $27{,}818$ / $13{,}909$ \\
element size (band / far)     & $h$              & ${\approx}\,\ell/2.4\ /\ {\approx}\,10^{-2}$ \\
\midrule
\multicolumn{3}{l}{\emph{Ductile testbed (AT1; reuses the elastic scale, $\mathcal{G}_c$, $\ell$ above)}}\\
degradation function          & rational (Wu)    & strength $\psi_c=1.5\times10^{-2}$ \\
yield stress / hardening exp.\ & $\sigma_y,\ n$  & $1.0$ / $5$ \\
reference plastic strain      & $\varepsilon_0$  & $5\times10^{-3}$ \\
mesh (diagonal band), elem / nodes & ---         & $24{,}463$ / $24{,}591$ \\
DOFs (displacement / damage)  & ---              & $49{,}182$ / $24{,}591$ \\
\bottomrule
\end{tabular}
\caption{Problem and discretization parameters for the mode-I testbed.}\label{tab:physics}
\end{table}

\begin{table}[htbp]
\centering
\begin{tabular}{lll}
\toprule
quantity & symbol & value \\
\midrule
\multicolumn{3}{l}{\emph{Outer trust region (Section~\ref{sec:tr})}}\\
initial radius              & $\Delta_0$              & $\norm{P^{-1}g}_P$ (field-split step scale) \\
maximum radius              & $\Delta_{\max}$         & $10^{8}$ \\
accept / expand ratio       & $\eta_1,\ \eta_2$       & $0.1,\ 0.75$ \\
shrink / expand factor      & ---                     & $0.25,\ 2.0$ \\
outer convergence on $\norm{R_\Iset}$ & atol, rtol    & $10^{-7},\ 10^{-6}$ \\
maximum outer iterations    & ---                     & $200$ \\
\midrule
\multicolumn{3}{l}{\emph{Steihaug--Toint CG (Section~\ref{sec:cg})}}\\
forcing tolerance           & $\varepsilon$           & $\min(0.1,\ \sqrt{\norm{g}/\norm{g_0}})$ \\
preconditioner              & $P$                     & symmetric block Gauss--Seidel \\
termination                 & ---                     & boundary / negative-curvature exit \\
\midrule
\multicolumn{3}{l}{\emph{Irreversibility PDAS (Section~\ref{sec:pdas})}}\\
dead-band                   & $\lambda_{\mathrm{tol}}$& $10^{-8}$ \\
bound tolerance             & $\varepsilon_{\mathrm b}$& $10^{-8}$ \\
multiplier scaling          & $B$                     & lumped mass on $d$ \\
\midrule
\multicolumn{3}{l}{\emph{Nonlinear elimination, NEPIN (Section~\ref{sec:nepin})}}\\
damage band thresholds      & $d_{\mathrm{lo}},\ d_{\mathrm{hi}}$ & $10^{-2},\ 10^{-2}$ \\
displacement front threshold& $\tau$                  & $10^{-2}$ \\
restricted assembly         & ---                     & on \\
\midrule
\multicolumn{3}{l}{\emph{Block sub-solves (field split / elimination)}}\\
type                        & ---                     & reduced-space VI Newton \\
linear preconditioner       & ---                     & algebraic multigrid (BoomerAMG) \\
Krylov / inner forcing      & ---                     & BiCGStab, Eisenstat--Walker \\
sub-solve convergence       & atol, rtol              & $10^{-8},\ 10^{-6}$ \\
\bottomrule
\end{tabular}
\caption{Algorithmic parameters. The bottom two rows apply to every field-split / elimination block
sub-solve.}\label{tab:params}
\end{table}

\paragraph{Robustness and physics} Table~\ref{tab:outcome} records which configurations complete the
loading history. Under the energy merit MONO, MSPIN, and NEPIN all finish the full $120$ steps with \emph{zero}
time-step cutbacks and produce \emph{identical} brittle physics in each regularization (peak reaction
$R_y=0.8127$ for AT2, $0.8744$ for AT1; $\max d\to1$, $\min d\approx0$) --- so the comparison is about
solver cost, not the answer (Table~\ref{tab:cost}). The residual
merit $\tfrac12\norm{R}^2$ fares far worse, as anticipated in Section~\ref{sec:intro}: monolithic TR-R
death-spirals to the time-step floor mid-propagation (step $68/120$, $21$ cutbacks), and although the
field-split sweep restores robustness, MSPIN-TR-R is impractically slow --- $2465$ outer iterations
reach only step $68$, more than twice MONO-TR-E's count for the \emph{entire} history. The residual
merit is therefore not a practical globalization here, and NEPIN is run under the energy merit
alone. For reference, the standard staggered (alternate-minimization) baseline also fails to complete. Its
fixed-point loop is iterated to a global relative/absolute residual tolerance of $10^{-6}$ with no imposed
sweep cap (and is \emph{not} accepted on a maximum-iteration count), so ``stagnation'' here is genuine slow
convergence rather than a truncation: near propagation the sweep count per step climbs into the many
hundreds ($\sim\!1000$), and by step $71/120$ the step can no longer meet tolerance before the time-step
floor $\mathrm{d}t_{\min}$ is reached, so the history does not complete. This is the slow-convergence mode
that motivates nonlinear preconditioning (Section~\ref{sec:intro}).

\begin{table}[htbp]
\centering
\begin{tabular}{l p{0.34\linewidth} p{0.40\linewidth}}
\toprule
solver & energy merit (TR-E) & residual merit (TR-R) \\
\midrule
MONO        & completes, $0$ cutbacks & death-spiral at step $68/120$ ($21$ cutbacks) \\
MSPIN       & completes, $0$ cutbacks & unfinished: $2465$ outer itrs reach step $68/120$ \\
NEPIN & completes, $0$ cutbacks & not pursued \\
\midrule
AM (staggered) & \multicolumn{2}{p{0.76\linewidth}}{stagnates at step $71/120$ (up to $\sim\!1000$ sweeps/step near propagation)} \\
\bottomrule
\end{tabular}
\caption{Robustness by globalization merit (AT2). The energy merit completes the full $120$-step history
for every family; the residual merit $\tfrac12\norm{R}^2$ does not (MONO death-spirals, MSPIN is
impractically slow), so it was not pursued for NEPIN. The staggered (alternate-minimization)
baseline, AM, is listed for reference and does not complete either.}\label{tab:outcome}
\end{table}

\subsection{Brittle fracture}\label{sec:res-brittle}

\begin{table}[htbp]
\centering
\begin{tabular}{lcc}
\toprule
metric & MONO & NEPIN \\
\midrule
\multicolumn{3}{l}{\emph{AT2} ($\alpha=d^2$, peak $R_y=0.8127$)}\\
outer (nonlinear) iterations           & 1100 & \textbf{795} \\
trust-region rejections                & 170  & 125 \\
Steihaug CG iterations ($A\!\cdot\!v$) & 2709 & 2227 \\
\midrule
\multicolumn{3}{l}{\emph{AT1} ($\alpha=d$, peak $R_y=0.8744$)}\\
outer (nonlinear) iterations           & 1011 & \textbf{753} \\
trust-region rejections                & 149  & 97 \\
Steihaug CG iterations ($A\!\cdot\!v$) & 2484 & 2105 \\
\bottomrule
\end{tabular}
\caption{Brittle fracture: cost of NEPIN vs.\ monolithic Newton over the full loading history (120 load
steps, energy merit). Both complete $120/120$ steps with $0$ cutbacks and identical physics (peak $R_y$
as listed; $\max d\to1$, $\min d\approx0$).}\label{tab:cost}
\end{table}

\begin{figure}[htbp]
\centering
\includegraphics[width=0.85\textwidth]{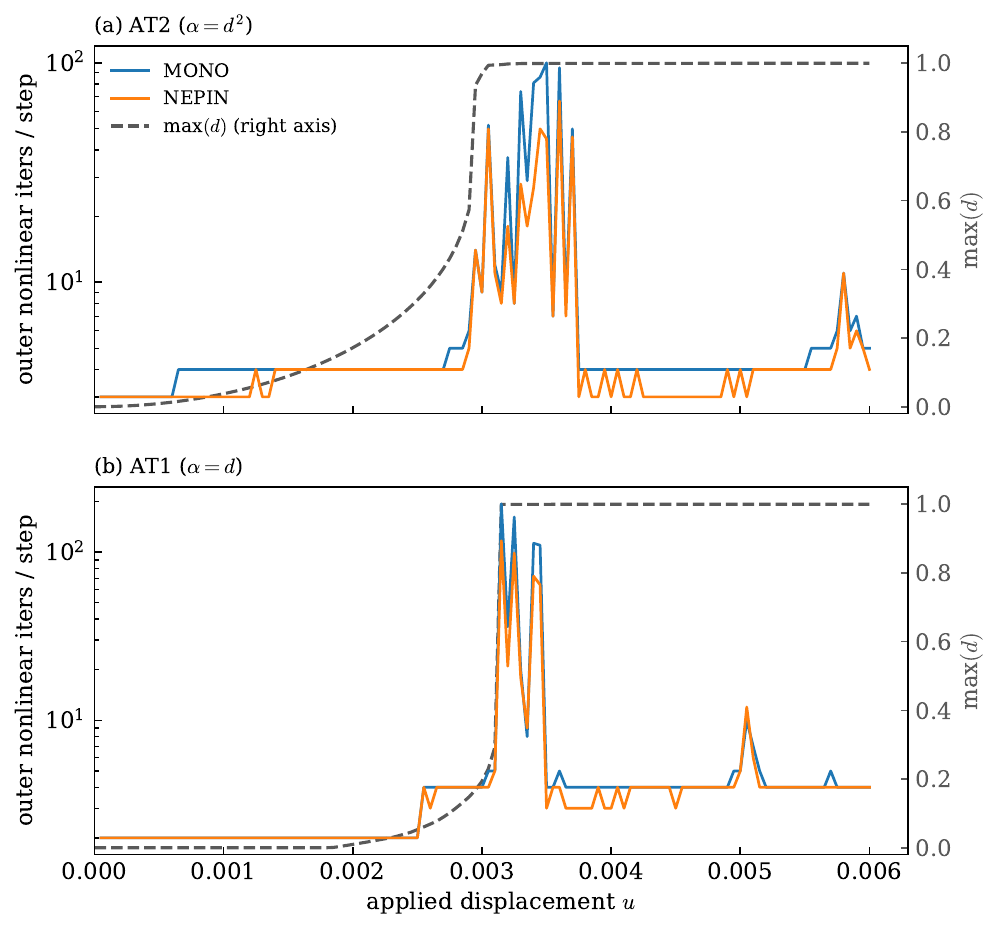}
\caption{Brittle fracture: outer nonlinear iterations per load step versus applied displacement $u$
(log scale, left axis), with $\max d$ overlaid (right axis), for (a)~the AT2 and (b)~the AT1
regularization. NEPIN and MONO are near-identical in the elastic regime and separate through the
nucleation/propagation spike, where NEPIN sits well below MONO. AT1's elastic threshold makes the spike
sharper and more localized than AT2's.}
\label{fig:iters}
\end{figure}

\paragraph{Iteration reduction} NEPIN cuts the outer (nonlinear) iteration count over monolithic Newton
by $28\%$ on AT2 ($1100\to795$) and $26\%$ on AT1 ($1011\to753$), at identical physics and zero
time-step cutbacks (Table~\ref{tab:cost}). The reduction concentrates where the nonlinearity does:
Figure~\ref{fig:iters} shows the per-step counts are near-identical in the elastic regime and separate
through the nucleation/propagation spike, where NEPIN tracks well below MONO. Eliminating the coupled
crack front removes most of the outer stall; how close this comes to the full-field sweep (MSPIN), and at what
block-work cost, is quantified in Section~\ref{sec:res-mspin}.

\paragraph{Assembly cost} With subdomain-restricted assembly (Section~\ref{sec:nepin}) the NEPIN
iterates are identical to full-domain assembly to numerical precision, so Table~\ref{tab:cost} is unchanged; the assembly
work per elimination scales with the hard set $\mathcal H$ instead of the mesh. How much that saves
depends on how localized $\mathcal H$ is, which is model-dependent (Table~\ref{tab:hardset}): AT2 has no
elastic threshold, so damage diffuses over most of the loaded domain and the damage band is broad, whereas
AT1's elastic threshold localizes damage and collapses the band roughly eightfold, so the restriction bites
far harder there. The displacement front is compact on both, with one exception --- at nucleation and near
crack-through its residual-based selection can momentarily span the whole mesh (Section~\ref{sec:discussion}). The saving is bounded, though: the restriction trims only the
\emph{elimination}'s assembly, whereas the outer full-mesh coupled residual and Jacobian --- which
dominate the per-iteration cost (next paragraph) and are independent of the hard set --- are untouched.
Enabling the restriction is thus the main assembly lever (on AT2 it more than halves the elimination's
assembly work); compacting the band further --- e.g.\ raising the damage-selection threshold to shrink
the band $\sim\!13\times$ (Section~\ref{sec:discussion}) --- sharpens locality but is close to a wash for
overall cost.

\begin{table}[htbp]
\centering
\begin{tabular}{lcc}
\toprule
hard set (\% of mesh DOFs) & AT2 & AT1 \\
\midrule
damage band $\mathcal H_d$: median (max)  & $38$ ($50$) & $4.6$ ($7.4$) \\
displacement front $\mathcal H_u$: median & $0.7$       & $0.6$ \\
\bottomrule
\end{tabular}
\caption{Hard-set size as a fraction of the mesh (energy-merit NEPIN, over the full loading history). The
damage band $\mathcal H_d$ collapses from broad (AT2, no elastic threshold) to compact (AT1), the
localization that makes the subdomain-restricted assembly bite harder on AT1. The displacement front
$\mathcal H_u$ is compact on both (median $\lesssim\!1\%$), except that its residual-based selection can
momentarily span the whole mesh at nucleation and crack-through.}\label{tab:hardset}
\end{table}

\paragraph{Cost structure} The per-iteration cost is dominated by \emph{assembly}, not linear algebra.
The coupled Hessian $A=\nabla^2\Psi$ --- both diagonal blocks and the off-diagonal coupling
$A_{01},A_{10}$ --- and the coupled residual are re-assembled every outer iteration, a cost \emph{shared}
by MONO and every NPC family; the Steihaug CG itself (matrix-free $A$-products, block-SGS preconditioner)
is a small fraction. NEPIN's \emph{additional} per-iteration cost over MONO is therefore narrow: the
block sub-solve sweep plus the residual evaluation at the swept iterate. That sweep is deliberately light
--- from Table~\ref{tab:cost}, NEPIN performs $\approx3$ block sub-solve iterations per outer
iteration (each on the hard set, with band-restricted assembly) versus $\approx5$ over the whole field
for the full-field sweep (Section~\ref{sec:res-mspin}), and none for MONO. Whether that trade is favorable
is not settled by iteration counts alone, since the full-domain outer work and the restricted sweep enter at
different element counts. We therefore reckon work with a machine-independent \emph{full-mesh-equivalent}
(FME) \emph{assembly-work proxy} $C_{\mathrm{asm}}$: each residual, Jacobian, or energy element-loop pass is
weighted by the number of elements it touches --- $N_e$ for a full-domain pass, $N_e(\mathcal H_k)$ for a
band-restricted sweep pass --- and summed over the run (including the extra passes from rejected
trust-region trials), then normalized by $N_e$. It counts element-loop passes, not their unequal per-pass
cost (a coupled Jacobian assembly is heavier than an energy evaluation), so it is a proxy for assembly work,
not a runtime surrogate. Section~\ref{sec:res-ondemand} reports $C_{\mathrm{asm}}$ for MONO and both NEPIN
variants; it shows the $26$--$28\%$ outer-iteration reduction does \emph{not} translate into total work
below monolithic Newton.

\subsection{Ductile fracture}\label{sec:res-ductile}

The ductile model --- small-strain $J_2$ plasticity whose plastic work drives damage through the
strongly-coupled E-P-PD formulation (\ref{app:models}) --- exercises the framework on a
dissipative problem. Its response is genuinely ductile (Figure~\ref{fig:load-disp}, right): an elastic
branch, a yield knee, a plastic-hardening stage, a peak reaction $R_y=0.544$, then a sharp post-peak
softening as the crack forms; and because the plastic work drives damage, the crack follows the
$\sim\!45^\circ$ plastic slip band from the notch tip (Figures~\ref{fig:crack},~\ref{fig:ep}, right)
rather than the straight mode-I path.

\begin{table}[htbp]
\centering
\begin{tabular}{lcc}
\toprule
metric (ductile AT1, E-P-PD; peak $R_y=0.544$) & MONO & NEPIN \\
\midrule
outer (nonlinear) iterations, total    & 1172 & \textbf{1052} \\
\quad at crack nucleation (one step)   & 199  & \textbf{121} \\
trust-region rejections                & 179  & 149 \\
Steihaug CG iterations ($A\!\cdot\!v$) & 3610 & 3262 \\
\bottomrule
\end{tabular}
\caption{Ductile fracture: cost of NEPIN vs.\ monolithic Newton (120 load steps, energy merit). Both
complete $120/120$ steps with $0$ cutbacks and identical physics ($\max d=0.73$).}\label{tab:ductile}
\end{table}

\begin{figure}[htbp]
\centering
\includegraphics[width=0.8\textwidth]{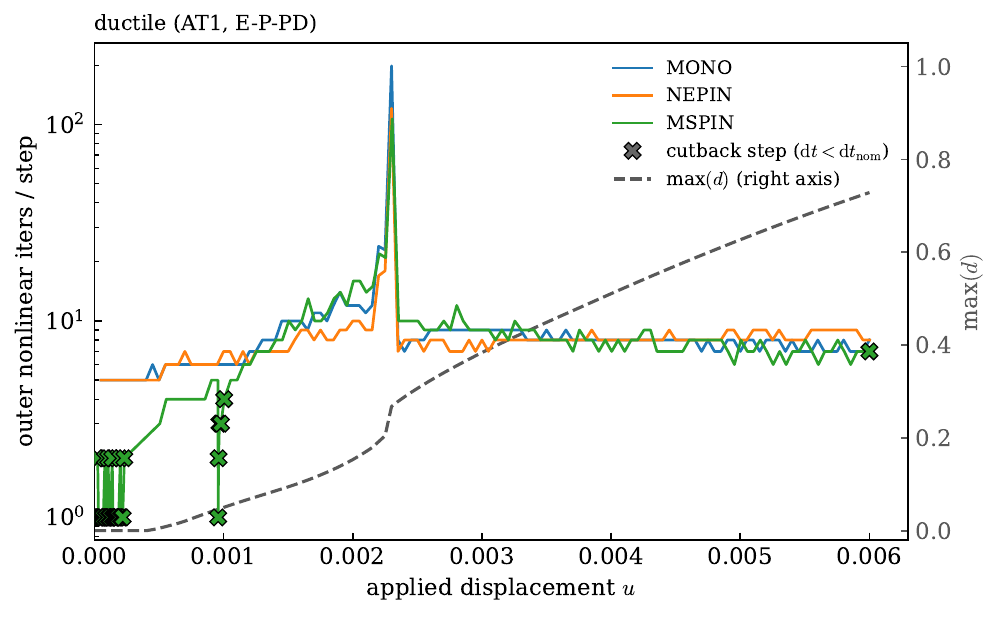}
\caption{Ductile fracture: outer nonlinear iterations per load step versus applied displacement $u$
(log scale), with $\max d$ overlaid. The cost concentrates in a single sharp spike at crack nucleation,
where NEPIN cuts monolithic Newton's $199$ outer iterations to $121$; elsewhere both are cheap, and both
complete with zero cutbacks. The full-field sweep (MSPIN) is overlaid for contrast: it is riddled with
time-step cutbacks ($\times$ markers, clustered in the elastic regime from the first load steps), the
fragility quantified in Section~\ref{sec:res-mspin}.}\label{fig:ductile-iters}
\end{figure}

Under the energy merit, both MONO and NEPIN complete the full history with zero cutbacks and identical
physics (Table~\ref{tab:ductile}). The outer-iteration cost is sharply concentrated at crack nucleation:
monolithic Newton spends $199$ outer iterations in the single nucleation step --- $17\%$ of its entire
budget --- and NEPIN cuts that spike to $121$ ($-39\%$), as Figure~\ref{fig:ductile-iters} shows. The
elastic and post-nucleation propagation steps are cheap ($\sim\!8$ iterations) for both, so the ductile
benefit is a concentrated cut at the hardest step ($-10\%$ overall) rather than the broad reduction of
the brittle case. Both the nonlinearity and NEPIN's advantage thus concentrate at the single nucleation
step, which localizes where any further gain would have to be found (Section~\ref{sec:discussion}).

\subsection{NEPIN vs.\ the full-field sweep (MSPIN)}\label{sec:res-mspin}

NEPIN restricts the multiplicative field-split sweep to the hard set; sweeping the \emph{whole} field is
MSPIN (Section~\ref{sec:nepin}). Two comparisons place NEPIN relative to it.

\paragraph{Brittle: NEPIN recovers most of the full-sweep reduction at less work} The full-field sweep
reaches the lowest outer-iteration counts (Table~\ref{tab:mspin}), but NEPIN recovers about $79\%$ (AT2)
and $99\%$ (AT1) of the MONO$\to$MSPIN reduction while performing fewer block sub-solve iterations
($2388$ vs.\ $3275$ on AT2): eliminating only the coupled crack front removes most of the stall at less
block work than sweeping the whole field.

\begin{table}[htbp]
\centering
\begin{tabular}{lcccc}
\toprule
outer iterations & MONO & MSPIN & NEPIN & NEPIN recovers \\
\midrule
AT2 & 1100 & 713 & 795 & $79\%$ \\
AT1 & 1011 & 750 & 753 & $99\%$ \\
\bottomrule
\end{tabular}
\caption{Brittle fracture: NEPIN recovers most of the full-field sweep's (MSPIN's) outer-iteration
reduction over monolithic Newton, at lower block-solve cost.}\label{tab:mspin}
\end{table}

\paragraph{Ductile: NEPIN is robust where the full sweep is not} On the harder ductile problem the
full-field sweep is not robust: its post-sweep iterate drives the outer trust region into repeated
rejections and time-step cutbacks from the very first (elastic) load step. It still limps to the end of
the history, but only through $35$ time-step cutbacks and $1355$ trust-region rejections --- inflating the
nominal $120$-step history to $144$ converged steps --- where monolithic Newton and NEPIN each complete in
$120$ steps with \emph{zero} cutbacks and an order of magnitude fewer rejections ($179$ and $149$;
Table~\ref{tab:ductile}). The cutback-riddled MSPIN curve in Figure~\ref{fig:ductile-iters} traces this,
its $\times$ markers clustered in the elastic regime. The fragility is specific to the full-field sweep:
MONO and NEPIN both complete cleanly on the same problem and mesh. The reason is specific: in the elastic
regime NEPIN's \emph{damage} hard set $\mathcal H_d$ is empty ($d\approx\dold$, below $d_{\mathrm{lo}}$), so
it sweeps only the displacement front $\mathcal H_u$ --- a benign elastic solve at fixed $d$ --- and never
drives the damage field prematurely; the full-field sweep instead updates $d$ over the whole domain and
pushes it onto the softening branch before the load warrants it, which the outer trust region then rejects
and cuts back. (The displacement front $\mathcal H_u$ is not empty --- the maximum-residual DOF always
qualifies --- but sweeping $u$ at fixed $d$ is harmless.) Once damage localizes, NEPIN's sweep is confined
to the crack front. The restricted sweep is thus not merely cheaper but more robust: it does not perturb
the fields where the coupled operator is sensitive.

To make the mechanism concrete, Table~\ref{tab:mspin-diag} traces the full-field sweep on the \emph{first}
(elastic) load step, where it already cuts back. Almost all damage DOFs are pinned at the irreversibility
bound ($24{,}510$ of $24{,}591$, $99.7\%$); the whole-field damage solve nonetheless moves the remaining
$\sim\!0.3\%$ inactive damage DOFs, and that alone is enough that the post-sweep iterate \emph{raises} the
coupled energy --- the trust-region ratio $\rho$ is negative at \emph{every} radius as $\Delta$ collapses
from $2.9\times10^{-3}$ to the floor, so no step is accepted and $\mathrm{d}t$ is cut. NEPIN never enters this state because its damage hard set is
empty here; MONO and NEPIN both complete this step in $\le4$ iterations.

\begin{table}[htbp]
\centering
\begin{tabular}{lccc}
\toprule
TR trial & radius $\Delta$ & ratio $\rho$ & step $\Pnorm{p}$ \\
\midrule
$1$ & $2.9\times10^{-3}$ & $-1.1\times10^{-3}$ & $1.48$ \\
$2$ & $7.2\times10^{-4}$ & $-1.6\times10^{-4}$ & $0.37$ \\
$3$ & $1.8\times10^{-4}$ & $-3.6\times10^{-5}$ & $0.092$ \\
$4$ & $4.5\times10^{-5}$ & $-8.8\times10^{-6}$ & $0.023$ \\
\multicolumn{4}{c}{$\vdots$\quad($\rho<0$ throughout as $\Delta$ collapses)} \\
$\to\Delta_{\min}$ & $1.1\times10^{-8}$ & $-3.3\times10^{-9}$ & $5.6\times10^{-6}$ \\
\bottomrule
\end{tabular}
\caption{Full-field sweep (MSPIN) failing on the \emph{first} ductile load step (elastic,
$t=5\times10^{-5}$): with $99.7\%$ of the damage DOFs pinned at the irreversibility bound, the whole-field
damage sweep still perturbs $d$, and the post-sweep step raises the energy, so the trust-region ratio
$\rho$ stays negative at every radius and the step is cut back. This is the mechanism behind the
$35$-cutback fragility; NEPIN (empty damage hard set here) and MONO complete the step in $\le4$
iterations.}\label{tab:mspin-diag}
\end{table}

\subsection{On-demand activation}\label{sec:res-ondemand}

Applied on every outer iteration, the sweep taxes the steps that do not need it
(Section~\ref{sec:nepin}); the on-demand gate confines it to the stalling steps. Table~\ref{tab:ondemand}
compares the gated (\emph{adaptive}) NEPIN with monolithic Newton and the always-on (\emph{blind}) NEPIN
across the three testbeds, using the fixed thresholds $\theta_{\mathrm{on}}=0.5$,
$\theta_{\mathrm{off}}=0.1$ (Section~\ref{sec:nepin}) --- not tuned per problem.

\begin{table}[htbp]
\centering
\begin{tabular}{lccccc}
\toprule
testbed & MONO & blind & adaptive & gate active & block its (blind$\,\to\,$adaptive) \\
\midrule
AT2     & 1100 & 795  & 828 & $42.5\%$ & $2388\to1429$ \\
AT1     & 1011 & 753  & 820 & $62.6\%$ & $2303\to2012$ \\
ductile & 1172 & 1052 & 967 & $75.2\%$ & $4514\to2242$ \\
\bottomrule
\end{tabular}
\caption{On-demand activation. Outer nonlinear iterations for monolithic Newton (MONO), always-on
(\emph{blind}) NEPIN, and gated (\emph{adaptive}) NEPIN; the fraction of outer iterations on which the
gate ran the sweep; and the sweep surcharge in block sub-solve iterations (blind~$\to$~adaptive). All
complete $120/120$ load steps with $0$ cutbacks; the gate uses the same thresholds on every
problem.}\label{tab:ondemand}
\end{table}

Three things follow. First, the activation fraction \emph{rises with problem difficulty} --- $42.5\%$
(AT2), $62.6\%$ (AT1), $75.2\%$ (ductile) --- so the gate spends the surcharge where the monolithic step
actually stalls (the nucleation and propagation regime) and skips it on the easy majority (the elastic
phase and the cheap post-crack steps), cutting the block sub-solve work by $13$--$50\%$ relative to blind
NEPIN. Second, the gate does not regress in these tests: it never exceeds the monolithic iteration count, and it
recovers most of blind NEPIN's reduction --- $89\%$ on AT2 and $74\%$ on AT1 --- at that reduced
surcharge. Third, on the ductile problem the gate is not merely cheaper but \emph{better} than blind NEPIN
($967$ versus $1052$ outer iterations): withholding the sweep on the easy steps avoids the same
counterproductive perturbations that make the full-field sweep fragile there
(Section~\ref{sec:res-mspin}), so the gated iterate path is also cleaner. The elimination is thus a safe
default --- it reduces to monolithic Newton where the step is healthy and engages only where it pays
\emph{in nonlinear iterations} (which, as Table~\ref{tab:casm} shows, is not always where it pays in total
assembly work).

\paragraph{Total assembly work} Cutting block work relative to blind NEPIN is only half the question; the
other half is whether the elimination is cheaper than \emph{monolithic Newton}. In the full-mesh-equivalent
count $C_{\mathrm{asm}}$ (Section~\ref{sec:res-brittle}) it is not (Table~\ref{tab:casm}): adaptive NEPIN
lands within $[-8\%,+10\%]$ of MONO across the three testbeds --- essentially cost-neutral --- while blind
NEPIN costs up to $+30\%$ (ductile), its surcharge outrunning its iteration saving, and the full-field sweep
(MSPIN) is $2.6$--$5.2\times$ MONO. What the on-demand gate buys is exactly this cost-competitiveness: it
holds the elimination within $\sim\!10\%$ of monolithic Newton where blind NEPIN does not --- a
bounded-overhead safeguard, though not literally cost-free (the ductile case is $+10\%$) --- but it does
\emph{not} make the method faster than MONO on these problems. The robustness of the overall solve is
supplied by the energy trust region, under which MONO itself completes every history without cutbacks; what
the elimination adds, once that globalization is fixed, is \emph{outer-iteration reduction} --- especially
the nucleation-spike reduction on the single hardest step --- at cost comparable to monolithic Newton, and
--- as Section~\ref{sec:res-mode2} confirms --- a cost that a local elimination does not push below MONO even
under mesh refinement.

\begin{table}[htbp]
\centering
\begin{tabular}{lccc}
\toprule
$C_{\mathrm{asm}}$ (FME) & MONO & blind NEPIN & adaptive NEPIN \\
\midrule
AT2     & $3640$ & $3854$\,($+6\%$)  & $3360$\,($-8\%$) \\
AT1     & $3331$ & $3271$\,($-2\%$)  & $3291$\,($-1\%$) \\
ductile & $3874$ & $5037$\,($+30\%$) & $4260$\,($+10\%$) \\
\bottomrule
\end{tabular}
\caption{Full-mesh-equivalent assembly-work proxy $C_{\mathrm{asm}}$ (Section~\ref{sec:res-brittle}), percentages
relative to MONO. Adaptive NEPIN is cost-competitive with monolithic Newton ($[-8\%,+10\%]$); blind NEPIN
can cost $+30\%$ (ductile) as its whole-history surcharge outruns the iteration saving; MSPIN (not shown)
is $2.6$--$5.2\times$ MONO. Once the energy trust region has supplied robustness, the elimination is an
iteration-reduction mechanism at $\sim$MONO assembly work, not a speedup.}\label{tab:casm}
\end{table}

\paragraph{Threshold sensitivity} The gate's thresholds are fixed, not tuned per problem;
Table~\ref{tab:m6} varies them one at a time on AT1. Over $\theta_{\mathrm{on}}\in[0.3,0.7]$ and
$\theta_{\mathrm{off}}\in[0.05,0.2]$ the outer-iteration count ($820$--$821$), activation fraction
($62.4$--$62.6\%$), and $C_{\mathrm{asm}}$ ($3290$--$3293$) are essentially unchanged. The reason is that
the reduced-residual contraction $\theta_k$ is sharply bimodal --- near $0$ while the coupled step
contracts, jumping to $\approx1$ when it stalls --- so any threshold in the wide gap between the two modes
classifies iterations identically. Although $\theta_k=\norm{R_{\Iset}(x_k)}/\norm{R_{\Iset}(x_{k-1})}$ is
dimensionless as a ratio, it is \emph{not} invariant to the relative scaling of the displacement and damage
blocks: rescaling one block reweights the mixed norm, so the ratio shifts whenever the two blocks' relative
contributions change between iterations. That the same thresholds transfer across the brittle and ductile
models is therefore an \emph{empirical} robustness property of these nondimensionalized testbeds --- the
observed $\theta_k$ is sharply bimodal, with converging and stalled iterations well separated, so thresholds
over a broad interval produce essentially identical classifications --- rather than a scale-invariance
guarantee; transferability under arbitrary rescaling would require a block-scaled or preconditioned norm.

\begin{table}[htbp]
\centering
\begin{tabular}{lccc}
\toprule
$(\theta_{\mathrm{on}},\theta_{\mathrm{off}})$ & outer iters & gate active & $C_{\mathrm{asm}}$ \\
\midrule
$(0.5,0.1)$ (base) & $820$ & $62.6\%$ & $3291$ \\
$(0.3,0.1)$        & $820$ & $62.6\%$ & $3291$ \\
$(0.7,0.1)$        & $821$ & $62.4\%$ & $3293$ \\
$(0.5,0.05)$       & $820$ & $62.6\%$ & $3291$ \\
$(0.5,0.2)$        & $820$ & $62.4\%$ & $3290$ \\
\bottomrule
\end{tabular}
\caption{On-demand threshold sensitivity (AT1): varying $\theta_{\mathrm{on}}$ and $\theta_{\mathrm{off}}$
one at a time about the fixed base $(0.5,0.1)$ leaves outer iterations, gate activation, and assembly work
essentially unchanged --- the bimodal contraction makes the gate insensitive to the threshold over a wide
range.}\label{tab:m6}
\end{table}

\subsection{Mesh sensitivity and a mode-II check}\label{sec:res-mode2}

Two questions remain: whether the elimination's advantage \emph{grows} under mesh refinement (a cost
crossover), and whether the algebraic hard-set selection depends on the mesh. Both are addressed on a
\emph{mode-II} (shear) single-edge-notched test --- a different geometry whose crack kinks and curves
down-right from the notch, so it is not aligned with any pre-refined band. The domain is the unit square
with a slit crack along the left half of the midline (tip at the centre), sheared on the top edge with the
bottom clamped; it is discretized by conforming, statically graded meshes whose \emph{lower-right quarter}
--- the region the crack enters --- is uniformly refined to $\ell/h\in\{1.2,2.4\}$ at fixed $\ell$
(broad-region refinement, not a path-conforming band, so the hard set cannot ``ride'' the refinement).
Unlike the nucleation-dominated mode-I tests, the pre-cracked mode-II specimen \emph{propagates} from the
notch, and monolithic Newton completes it benignly (peak $15$ outer iterations at crack-through, $0$
cutbacks).

\begin{table}[htbp]
\centering
\begin{tabular}{lcccc}
\toprule
$\ell/h$ ($N_e$) & MONO & blind NEPIN & NEPIN cuts & $C_{\mathrm{asm}}$ MONO\,/\,NEPIN \\
\midrule
$1.2$ ($4312$) & $1415$ & $1341$ & $5.2\%$ & $4275$ / $6983$ \\
$2.4$ ($8208$) & $1483$ & $1413$ & $4.7\%$ & $4529$ / $7203$ \\
\bottomrule
\end{tabular}
\caption{Mode-II mesh sensitivity: outer nonlinear iterations and full-mesh-equivalent assembly-work proxy
$C_{\mathrm{asm}}$ for monolithic Newton and always-on (blind) NEPIN at two refinements of the lower-right
quarter, fixed $\ell$. Both complete with $0$ cutbacks. Adaptive NEPIN is not tabulated here; on this benign
problem it activates on only $\sim\!9\%$ of iterations and tracks MONO in outer iterations
($1410$ vs.\ $1415$ at $\ell/h=1.2$; see text).}\label{tab:mode2}
\end{table}

Three observations. First, \emph{no crossover}: NEPIN's iteration advantage is a flat $\sim\!5\%$ at both
resolutions --- the gap does not widen as $h$ falls --- and its $C_{\mathrm{asm}}$ stays $\sim\!60\%$
\emph{above} monolithic Newton, the whole-field sweep surcharge again outweighing the small iteration
saving. The mode-II difficulty does not localize with refinement; it remains the shared limit-point cost of
crack-through, which a local elimination cannot reduce (Section~\ref{sec:discussion}). Second, the gate
behaves as designed on this benign problem: it activates on only $\sim\!9\%$ of iterations, and adaptive
NEPIN closely tracks monolithic Newton in outer iteration count ($1410$ vs.\ $1415$ at $\ell/h=1.2$). Third, the \emph{hard-set selection is mesh-stable}: the residual-selected displacement front
$\mathcal H_u$ is a compact $3$--$5\%$ of the mesh (p90 $6$--$9\%$) that if anything \emph{shrinks} with
refinement, as a fixed physical width should, so the raw-residual criterion in~\eqref{eq:hardset} is not
biased by the graded element sizes; the damage band $\mathcal H_d$ is broad ($\sim\!50\%$ on this AT2
problem) but likewise mesh-independent. That broad $\mathcal H_d$ is the main target for a tighter,
gradient-based selection (Section~\ref{sec:discussion}).

\section{Discussion}\label{sec:discussion}

\paragraph{Energy merit} Every family completes with zero cutbacks under the energy merit, while the
residual merit $\tfrac12\norm{R}^2$ does not (Table~\ref{tab:outcome}); $\Psi$ is the natural,
well-conditioned merit for this minimization, and the decoupled acceptance~\eqref{eq:accept} is what
keeps it from stalling near the constrained minimizer.

\paragraph{Robustness and on-demand cost} Across brittle and ductile fracture the coupled-front
elimination is \emph{robust}: it completes every history with zero cutbacks, where monolithic
residual-merit Newton death-spirals, alternate minimization stalls, and, on the ductile problem, even
the full-field sweep loses robustness (Section~\ref{sec:res-mspin}). Where monolithic Newton itself
converges, the elimination reduces outer iterations by $26$--$28\%$ (brittle) and cuts the ductile
single-step nucleation spike by about $40\%$. The sweep is not free, however: each application adds the
field-split block sub-solves --- a handful of extra residual/Jacobian assemblies per outer iteration
(Section~\ref{sec:res-brittle}). Applied on \emph{every} iteration this surcharge is largely wasted,
since monolithic Newton already converges in $\approx\!4$ iterations on the great majority of the brittle
steps and only the crack-front steps repay the sweep. This is exactly what the \emph{on-demand} gate
(Section~\ref{sec:nepin}) addresses: it holds the sweep off while the monolithic step is contracting and
engages it only where the step stalls, so the elimination adds no surcharge on the easy majority of steps
and reduces to monolithic Newton there, while retaining the nucleation-spike reduction where it is needed
(quantified in Section~\ref{sec:res-ondemand}). In the machine-independent assembly-work proxy
$C_{\mathrm{asm}}$, adaptive NEPIN is cost-competitive with monolithic Newton (within
$[-8\%,+10\%]$) while blind NEPIN costs up to $+30\%$: the gate's role is to \emph{secure} that
cost-competitiveness, not to produce a speedup. We therefore claim no cost advantage over monolithic
Newton. The robustness here is supplied by the energy trust region --- under it, MONO alone completes every
history without cutbacks; once that globalization is fixed, what the elimination adds is outer-iteration and
nucleation-spike reduction at comparable cost, and Section~\ref{sec:res-mode2} finds this unchanged under
mesh refinement (no crossover). That a purely \emph{local} elimination does not beat monolithic Newton here
is itself informative: the dominant expense is the limit-point (softening) regime --- a global instability
the trust region already handles --- which a hard-set correction cannot reduce below MONO. This also settles
the question raised in Section~\ref{sec:intro}: once the globalization is fixed at the energy trust region
and cost is measured machine-independently against a well-globalized monolithic Newton, the preconditioner's
benefit is outer-iteration reduction, not wall-clock speed. The speedups reported against alternate
minimization~\citep{KopanicakovaKothariKrause2023} therefore cannot be attributed to nonlinear
preconditioning alone; the present results show that, on our benchmarks, once the globalization is fixed at
the energy trust region and the comparison is made against a well-globalized monolithic Newton, the
nonlinear preconditioner reduces nonlinear iteration counts but not the overall assembly-work proxy.

\paragraph{Hard-set localization and selection} The restriction's benefit scales with how localized
$\mathcal H$ is, which the results make concrete: the damage band is broad on AT2 ($\sim\!38\%$ of the
mesh) but compact on AT1 ($\sim\!5\%$), tracking the model's elastic threshold. The selection metric is a
further lever --- raising the damage threshold from $d_{\mathrm{lo}}=10^{-2}$ to $d_{\mathrm{lo}}=0.25$ on AT2 shrinks the
damage band about $13\times$ (median $38\%\to3\%$) with a negligible outer-iteration penalty
($795\to802$) and identical physics: the strongly-damaged core carries the nonlinearity while the diffuse
tail can be left to the far field. The gain there is locality, not speed --- as noted under total assembly work,
the band-independent full-mesh coupled assembly dominates, so a tighter band barely changes overall work. The displacement front, selected by residual magnitude, is compact
through most of the history but can momentarily span the whole mesh at nucleation and crack-through,
which is where the residual criterion is least selective; keeping that front localized without
compromising robustness is the main remaining lever on the peak iteration count.

\paragraph{Dissipative (ductile) extension} The framework requires only that each load step is the
minimization of a potential whose stationarity is the coupled residual --- a structure that survives
\emph{dissipative}, elasto-plastic bulk response. For rate-independent $J_2$ plasticity the per-step
\emph{incremental} potential adds a stored plastic energy $g_p(d)\,\psi_p(\bar\varepsilon^{p})$ to
$\Psi$~\eqref{eq:psi-form}, with the constitutive return mapping as its inner minimizer (variational
plasticity~\citep{OrtizStainier1999}); consequently $R=\nabla\Psi$ and the $2\times2$ block structure
of $A$ carry over exactly. The identical solver --- energy trust region, PDAS, and field-split /
nonlinear-elimination sweep --- therefore applies with no modification to a strongly-coupled ductile
(E-P-PD) phase-field model~\citep{HuEtAl2021}. Section~\ref{sec:res-ductile} bears this out: both MONO and
NEPIN complete with identical physics, and NEPIN stays robust and cuts the crack-nucleation cost where the
full-field sweep loses robustness (Section~\ref{sec:res-mspin}). The constitutive form and incremental
variational structure are given in \ref{app:models}.

\section{Conclusions and future work}\label{sec:conclusions}
This work developed a nonlinear-elimination preconditioned trust-region Newton solver for
bound-constrained phase-field fracture: an energy-merit Steihaug--Toint trust region wrapped in a
primal--dual active set for irreversibility, with a merit-safeguarded field-split sweep that eliminates
an algebraically-identified, coupled crack-front hard set spanning both fields. Across brittle (AT1, AT2)
and strongly-coupled ductile ($J_2$ E-P-PD) mode-I benchmarks the robustness is supplied by the energy
trust region: under it, monolithic Newton alone completes every loading history with zero time-step
cutbacks, where residual-merit Newton death-spirals, alternate minimization stalls, and the full-field
sweep loses robustness (severe trust-region rejection and cutbacks) on the ductile problem. With that
globalization fixed, the \emph{on-demand} elimination reduces outer nonlinear iterations by $19$--$25\%$
over monolithic Newton on the brittle benchmarks and $17\%$ overall on the ductile one --- always-on
elimination reaches $26$--$28\%$ (brittle) and about $39\%$ at the ductile nucleation step, of which the
gate recovers most of the brittle reduction and improves the ductile whole-history count --- at identical
physics. Because the sweep's surcharge is repaid only where the monolithic
step stalls, the elimination is applied \emph{on demand}, gated by that step's own convergence: it engages
on $43$--$75\%$ of iterations, never regresses below monolithic Newton in iteration count, and on the
ductile problem even improves on the always-on elimination. Measured in a machine-independent
full-mesh-equivalent assembly-work proxy, however, the method is cost-\emph{competitive} with monolithic
Newton, not faster --- within $\sim\!10\%$, where an always-on sweep costs up to $+30\%$ --- and this does
not change under mesh refinement (Section~\ref{sec:res-mode2}). Its contribution is therefore an
iteration-reduction mechanism whose overhead the on-demand gate bounds, preserving the
energy-trust-region method's robustness without a demonstrated total-work advantage over it.

Why a purely \emph{local} elimination does not beat monolithic Newton here points to the future work: the
bottleneck is the softening \emph{limit point} --- a global instability the trust region already handles ---
rather than a local imbalance a hard-set correction removes, and the present sweep advances the crack front
only a sliver per iteration because it sweeps the coupled hard set just once. Three directions follow.
(i)~\emph{Converge} the elimination on a \emph{tightly confined} hard set --- for instance interaction
elements selected by the damage-gradient norm $\lVert\nabla d\rVert$, which makes the displacement and
damage hard sets coincide geometrically so the coupled $2\times2$ tangent can be solved together --- so the
front is resolved in one outer step at a surcharge small enough for the iteration saving to become a genuine
cost reduction. (ii)~Add a \emph{global}-mode correction (deflation of the crack-opening near-null-vector,
or continuation) to attack the limit point the local elimination cannot. (iii)~Since the construction needs
only a variational structure, couple it to a single-potential phase-field model with a genuine strength
surface (Section~\ref{sec:intro}), extending the solver to strength-driven nucleation.

\section*{Declaration of generative-AI use}
In preparing this work, the author used Anthropic's Claude models --- Opus~4.8, Sonnet~5, and Opus~5 ---
to assist with the software implementation and with setting up the comparison harness. All AI-assisted
output was reviewed and verified by the author, who takes full responsibility for the content and the
results presented in this manuscript.

\appendix
\section{Governing equations and constitutive models}\label{app:models}

This appendix specifies the boundary-value problem and the constitutive models that realize the energy
$\Psi$ of \eqref{eq:psi-form}; the solver of Sections~\ref{sec:tr}--\ref{sec:full} depends only on its
variational structure, not on these choices.

\paragraph{Strong form} Stationarity of $\Psi$ over $x=(u,d)$ subject to $\dold\le d\le1$ gives, at
each load step, quasi-static momentum balance coupled to a gradient-damage (Allen--Cahn-type)
evolution:
\begin{align}
  \nabla\!\cdot\bm\sigma &= \bm 0,\qquad \bm\sigma=\partial_{\bm\varepsilon}\psi , \label{eq:app-mom}\\
  \dfrac{\mathcal{G}_c}{c_0\,\ell}\Big(\alpha'(d)-2\ell^2\,\Delta d\Big) + g'(d)\,H &\;\ge\;0,
    \qquad \dold\le d\le1 , \label{eq:app-pff}
\end{align}
with $\bm\sigma$ the degraded stress and $H=\psi_e^{+}+\psi_p$ the crack driving energy.
Equation~\eqref{eq:app-pff} holds as a variational inequality (complementarity with the irreversibility
bound), enforced by the primal--dual active set of Section~\ref{sec:pdas}. The boundary conditions are
those of the mode-I single-edge-notched tension test of Section~\ref{sec:results}
(Table~\ref{tab:physics}).

\paragraph{Elasticity, degradation, and fracture (brittle)} The bulk is small-strain isotropic
elasticity with a spectral tension/compression split~\citep{MieheWelschingerHofacker2010}: $\psi_e^{+}$
and $\psi_e^{-}$ are the tensile (active, degraded) and compressive (inactive) parts of the elastic
strain energy. The crack geometric function and its normalization
$c_0=4\int_0^1\!\sqrt{\alpha(s)}\,\mathrm ds$ are
\begin{equation}
  \text{AT2: }\ \alpha(d)=d^2,\ c_0=2; \qquad \text{AT1: }\ \alpha(d)=d,\ c_0=\tfrac{8}{3}
\end{equation}
\citep{PhamAmorMarigoMaurini2011}. Two degradation functions are used: the quadratic
$g(d)=(1-d)^2(1-\eta)+\eta$ (the brittle testbeds), and the length-insensitive rational (Lorentz/Wu)
form
\begin{equation}
  g(d)=\dfrac{(1-d)^p}{(1-d)^p+a_1\,d\,(1+a_2 d)}\,(1-\eta)+\eta ,
  \qquad a_1=\dfrac{\mathcal{G}_c}{\psi_c}\,\dfrac{\xi}{c_0\,\ell},\ \ \xi=\alpha'(0),
\end{equation}
whose strength parameter $\psi_c$ fixes the peak stress independently of $\ell$~\citep{Wu2017} (the
ductile testbed); $\eta$ is a small residual-stiffness floor.

\paragraph{Plasticity (ductile)} The ductile testbed augments the bulk with small-strain $J_2$ (von
Mises) plasticity: an additive split $\bm\varepsilon=\bm\varepsilon_e+\bm\varepsilon_p$, the yield
function $\phi=\lVert\mathrm{dev}\,\bm\sigma\rVert-\sqrt{\tfrac23}\,g_p(d)\,\sigma_Y(\bar\varepsilon^{p})$
with associative flow, and the stored plastic energy
\begin{equation}
  \psi_p(\bar\varepsilon^{p})=\dfrac{n}{n+1}\,\sigma_y\,\varepsilon_0
    \Big[(1+\bar\varepsilon^{p}/\varepsilon_0)^{(n+1)/n}-1\Big],
  \qquad \sigma_Y(\bar\varepsilon^{p})=\dfrac{\partial\psi_p}{\partial\bar\varepsilon^{p}}
    =\sigma_y\,(1+\bar\varepsilon^{p}/\varepsilon_0)^{1/n}
\end{equation}
(power-law hardening; $\sigma_Y$ is the flow stress, with $\sigma_y$, $n$, $\varepsilon_0$ the yield
stress, exponent, and reference plastic strain). We adopt the strongly-coupled \emph{E-P-PD} model of
Hu et al.~\citep{HuEtAl2021}: $g_p=g$, so the phase field degrades the yield surface and the plastic
energy contributes to the crack driving force $H=\psi_e^{+}+\psi_p$. The plastic response enters through local internal variables $\alpha=(\bm\varepsilon_p,\bar\varepsilon^{p})$
carried at the quadrature points. Following the incremental variational formulation of
plasticity~\citep{OrtizStainier1999}, the per-step incremental potential
$\Pi_{\Delta t}(u,d,\alpha;\alpha_n)$ is minimized \emph{locally} over $\alpha$ (with $u,d$ held fixed and
$\alpha_n$ the converged previous-step values), which defines the \emph{reduced} incremental potential
\begin{equation}\label{eq:reduced-potential}
  \widehat\Psi(u,d)=\min_{\alpha}\ \Pi_{\Delta t}(u,d,\alpha;\alpha_n) .
\end{equation}
The radial return mapping performs exactly this local minimization, returning $\alpha^\star(u,d)$. By the
envelope theorem the partial derivatives with respect to $\alpha$ vanish at $\alpha^\star$, so the global
residual is the total derivative $R=\nabla_{(u,d)}\widehat\Psi$ and the coupled tangent
$A=\nabla^2_{(u,d)}\widehat\Psi$ is delivered by the algorithmically consistent constitutive tangent.
Consequently $A$ is symmetric and the $2\times2$ block structure holds for the ductile problem exactly as
in the brittle case (the $\widehat\Psi$ here plays the role of $\Psi$ throughout), so the solver applies
without modification.

\section{Implementation}\label{app:impl}

The solver is implemented as a custom coupled executor in the MOOSE finite-element
framework~\citep{PermannEtAl2020} driving PETSc~\citep{PETScUsersManual2026,BalayEtAl1997PETSc}, with the phase-field
kernels/materials in an application layer. The two fields are separate MOOSE nonlinear systems
(displacement, damage); the coupled operator $A$ is a PETSc \texttt{MatNest} of the per-system
matrices and their off-diagonal coupling blocks.

\paragraph{Sub-solves} Each field-split / elimination block sub-solve is a reduced-space
variational-inequality Newton solve (PETSc \texttt{SNESVINEWTONRSLS})~\citep{BensonMunson2006}
preconditioned by algebraic multigrid (hypre BoomerAMG~\citep{FalgoutYang2002,HensonYang2002}), with
Eisenstat--Walker inexact linear tolerances. The reduced-space solver updates only the inactive (free)
reduced space, so the linear solve is already free-set-sized; absent the assembly restriction, only the
finite-element assembly is full-domain. Active DOFs are held fixed --- their lower and upper bounds set
equal to the current value --- so the inner solve reuses the outer active set. Where a direct solve is
wanted, the reduced (band) coupled solves and block factorizations use the sparse direct solver
MUMPS~\citep{AmestoyEtAl2001MUMPS,AmestoyEtAl2006MUMPS}. The subdomain-restricted assembly limits the element loop to the locally-owned
elements incident to $\mathcal H$, reset after each sub-solve.

\bibliographystyle{unsrtnat}
\bibliography{ref}

\end{document}